\documentclass[english]{article}
\usepackage[T1]{fontenc}
\usepackage[utf8]{inputenc}
\usepackage{geometry}
\usepackage{babel}
\usepackage{float}
\usepackage{textcomp}
\usepackage{amsmath}
\usepackage{amsthm}
\usepackage{graphicx}
\usepackage{setspace}
\usepackage[unicode=true,pdfusetitle,
 bookmarks=true,bookmarksnumbered=false,bookmarksopen=false,
 breaklinks=false,pdfborder={0 0 1},backref=false,colorlinks=false]
 {hyperref}

\makeatletter
\addto\captionsenglish{%
}

\usepackage{fancyhdr}
\makeatother

\begin{document}
\begin{doublespace}
\begin{center}
\textbf{\Large{}A Tale of Two Currencies: Cash and Crypto}{\Large\par}
\par\end{center}

\begin{center}
\textbf{\large{}Ravi Kashyap (ravi.kashyap@stern.nyu.edu)}\footnote{\begin{doublespace}
Numerous seminar participants, particularly at a few meetings of the
econometric society and various finance organizations, provided suggestions
to improve the paper. The following individuals have been a constant
source of inputs and encouragement: Joshua Hong and the team at Formation
Fi; Dr. Yong Wang, Dr. Isabel Yan, Dr. Vikas Kakkar, Dr. Fred Kwan,
Dr. Costel Daniel Andonie, Dr. Guangwu Liu, Dr. Jeff Hong, Dr. Humphrey
Tung and Dr. Xu Han at the City University of Hong Kong. The views
and opinions expressed in this article, along with any mistakes, are
mine alone and do not necessarily reflect the official policy or position
of either of my affiliations or any other agency.
\end{doublespace}
}\textbf{ }
\par\end{center}

\begin{center}
\textbf{\large{}Estonian Business School / City University of Hong
Kong}{\large\par}
\par\end{center}

\begin{center}
\begin{center}
\today
\par\end{center}
\par\end{center}

\begin{center}
\par\end{center}

\begin{center}
Journal of Economic Literature Codes: D7: Analysis of Collective Decision-Making;
D8: Information, Knowledge, and Uncertainty; I31: General Welfare,
Well-Being; O3 Innovation ,Research and Development, Technological
Change, Intellectual Property Rights; XYZ: Creation of Universal Identities
(New JEL Code to be Added)
\par\end{center}

\begin{center}
studies; XYZ: Mathematical Techniques for Creating Universal Identities
(New MSC Code to be Added)
\par\end{center}

\begin{center}
\newpage{}
\par\end{center}

\begin{center}
\tableofcontents{}\newpage{}
\par\end{center}
\end{doublespace}

\section{Abstract }

We discuss numerous justifications for why crypto-currencies would
be highly conducive for the smooth functioning of today's society.
We provide several comparisons between cryptocurrencies issued by
blockchain projects, crypto, and conventional government issued currencies,
cash or fiat. We summarize seven fundamental innovations that would
be required for participants to have greater confidence in decentralized
finance (DeFi) and to obtain wealth appreciation coupled with better
risk management. 

The conceptual ideas we discuss outline an approach to: 1) Strengthened
Security Blueprint; 2) Rebalancing and Trade Execution Suited for
Blockchain Nuances 3) Volatility and Variance Adjusted Weight Calculation
4) Accommodating Investor Preferences and Risk Parity Construction;
5) Profit Sharing and Investor Protection; 6) Concentration Risk Indicator
and Performance Metrics; 7) Multi-chain expansion and Select Strategic
Initiatives including the notion of a Decentralized Autonomous Organization
(DAO).

Incorporating these concepts into several projects would also facilitate
the growth of the overall blockchain eco-system so that this technology
can, have wider mainstream adoption and, fulfill its potential in
transforming all aspects of human interactions.
\begin{doublespace}
\begin{center}
\newpage{}
\par\end{center}
\end{doublespace}

\section{\label{sec:MMT-and-MPT}MMT and MPT are Starting to Sound Empty }

There is a debate raging amongst economists and politicians that goes
to the very heart of what governments should and shouldn’t do to manage
future prosperity. The monetary and fiscal policies adopted by many
nations, over the last few decades, have garnered strong support for
the so called Modern Monetary Theory (MMT) (Mankiw 2020; Wray 2015).
MMT’s proponents claim that any nation that produces its own sovereign
currency (fiat or cash) cannot run out of money because it can always
just print more. In other words, the government essentially has no
financial constraints. 

MMT was originally a description of how spending in the economy already
happens. In that sense, the debate isn’t so much whether it should
or shouldn’t be implemented, but to what degree and under what circumstances.
Challengers say MMT would be highly irresponsible mismanagement of
the economy. The policies, they say, will lead to a massive increase
of the money supply that is bound to trigger inflation at levels not
seen since the seventies and eighties and perhaps even trend higher.
The application of MMT will require tax increases, to control any
inflationary pressures, which can be hugely unpopular and hard to
implement. 

The debate boils down to whether we believe that politicians and officials
have the data, knowledge and skills to delicately balance their spending
to deliver full employment while hitting an inflation target. MMT,
coming at a time of general economic uncertainty, could have a profound
impact on investment management, Decentralized Finance (DeFi: Zetzsche,
Arner \& Buckley 2020; End-note \ref{enu:Decentralized-finance})
and how corporations and ordinary citizens secure their wealth. 

Modern Portfolio Theory (MPT) is the theory behind most current financial
investment strategies. It uses elegant mathematics to formalize many
intuitive ideas about risk and return. MPT is one of the primary tools
used by fund managers to construct portfolios that match expected
reward to accepted risk (Elton \& Gruber 1997; Goetzmann, et al. 2014;
Fabozzi, Gupta \& Markowitz 2002). MPT is driven by a wide diversification
of assets to even out any downturns and achieve consistent growth.
Asset classes such as bonds are usually included to reduce risk, But
when the US Treasury bond yields dip to low levels, such as the recent
drop of 10 year US Treasury bond yields to below 1\% as it happened
during 2020, investors are having to look around for alternatives.
Bitcoin is suddenly coming up in more financial planning conversations
(End-note \ref{enu:United-States-Department-Bond-Yields}). 

\begin{figure}[H]
\includegraphics[width=12cm]{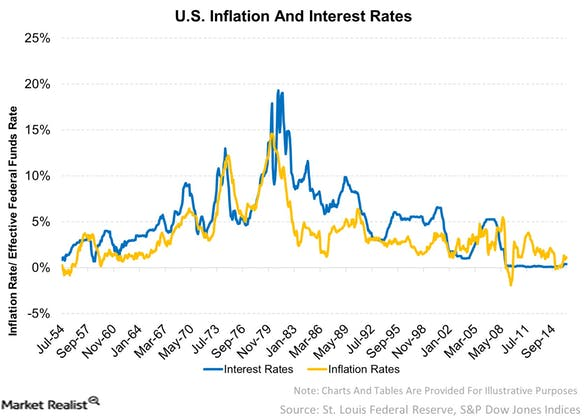}

\caption{\label{fig:Interest-Rates-Inflation}Interest Rates and Inflation}

Source: St. Louis Federal Reserve, S\&P Dow Jones Indices
\end{figure}

Despite its success, MPT is under threat (Miccolis 2012; Curtis 2002;2004).
MPT is not immune from inflation. Economic expectations are priced
in, but unexpected economic shocks are not. An economy running according
to MMT principles carries a higher risk of missing the inflation target
as the government tries to juggle its spending and taxation. Bonds
offer no protection here either, as witnessed during the high inflation
1970s, when the spread between bond yields and inflation converged
significantly compared to previous years (Laidler \& Parkin 1975;
Blinder 1982; DeLong 1997; Boschen \& Weise 2003; Figure \ref{fig:Interest-Rates-Inflation}). 

Another environmental risk comes from economic growth targets. MMT’s
main aim is to secure full employment thereby maximizing productivity
and GDP. Again, as governments try to spend and tax their way to their
objectives, they are unlikely to get consistent results (Palley 2015;
Epstein 2019; Baker \& Murphy 2020). 

\begin{doublespace}

\end{doublespace}

\subsection{\label{subsec:Outline-of-the}Outline of the Sections Arranged Inline}

Section (\ref{sec:MMT-and-MPT}), which we have already seen, provides
an introductory overview of the monetary policies currently being
pursued by most governments and the wealth management strategies commonly
used by many traditional financial firms. Section (\ref{sec:Interest-in-Interest}),
develops an analogy that helps us to understand the role of money
in society and then to see how crypto can fulfill the functions of
money, as we know it, in a better way. Sections (\ref{sec:Since-Bitcoin-Was-Coined};
\ref{sec:Decrypting-Crypto-and}) look at the origins of crypto-currencies
and the evolution of decentralized finance. Section (\ref{sec:Back-To-The-Future-Decentralized-Centralized-Back})
looks at how money has moved from being centralized to decentralized
and how there are attempts to make it centralized again using blockchain
technology. Section (\ref{sec:Crypto-Conundrums-versus}) is a discussion
of how crypto can be a remedy to several monetary maladies and the
promise it holds for creating equal wealth generation opportunities
for everyone. Section (\ref{sec:Bringing-Risk-Parity}) summarizes
several innovations that would be necessary to make blockchain wealth
management accessible and safe for the masses. 

The seven sub-sections in Section (\ref{sec:Bringing-Risk-Parity})
cover: 1) Strengthened Security Blueprint (Section \ref{sec:DeFi-Security:-Turning});
2) Rebalancing and Trade Execution Suited for Blockchain Nuances (Section
\ref{sec:Trade-Execution:-To}); 3) Volatility and Variance Adjusted
Weight Calculation (Section \ref{sec:VVV-Weight-Calculations:});
4) Accommodating Investor Preferences and Risk Parity Construction
(Section \ref{sec:The-Risk-Parity}); 5) Profit Sharing and Investor
Protection (Section \ref{sec:Caring-for-the-Community}); 6) Concentration
Risk Indicator and Performance Metrics (Section \ref{sec:Raising-the-Bar-CRI});
and 7) Multi-chain expansion and Select Strategic Initiatives (Section
\ref{sec:Multichain-Expansion-Select-Strategic}). This present article
provides a conceptual overview of these topics since these ideas will
be published in seven separate articles, with detailed mathematical
formulations and software architectural design considerations where
applicable.

\section{\label{sec:Interest-in-Interest}Interest in Interest Rates, Inflation
and Money Machines!!! }

We look at a simple analogy (Kashyap 2015) to get a better understanding
of money, interest rates, inflation and the importance of these concepts
for economic growth and wealth generation including how technology
is shaping the future of money. Sweeney \& Sweeney (1977) is a very
interesting tale of interest rates and inflation. Cochrane (2009)
has a discussion of money machines as they are understood in finance
(End-note \ref{enu:Money-Machines}).

Water gives life and sustains it (Franks 2000; Chaplin 2001; Westall
\& Brack 2018). It is required everywhere for life, as we know it,
to exist. In a similar vein, it is hard to imagine a modern economy
without money or money-equivalents. This comparison is only partly
valid since life, as we know it, would cease to prevail without water.
While we can essentially have a barter economy without money-equivalents.
Barring this key limitation, the smooth functioning of a modern economy
requires the flow of money-equivalents. 

Money has three main utilities: it serves as a medium of exchange,
a unit of measurement and a storehouse for wealth (Brunner \& Meltzer
1971; McLeay, Radia \& Thomas 2014). Water has numerous uses, but
we list three main ones to develop our comparison: it helps to transports
nutrients and minerals both within our bodies and all around us; it
regulates the temperature of our bodies and the external environment
and gives shape and structure to many things around us; and it dissolves
nutrients and stores them, more than any other substance known to
mankind. 

We have constructed elaborate devices and machines, to control and
divert the flow of water, to maximize the growth of life (Rogers \&
Fiering 1986; Pahl-Wostl 2008; Cosgrove \& Loucks 2015). Likewise,
we have the financial services sector, that controls and diverts the
flow of money-equivalents, to maximize the growth of an economy. Taking
the analogy a step further: our central reservoirs, irrigation canals,
water tankers, pumping stations, pipes and water sprinklers are devised
to keep water flowing around. Similarly, centralized and regional
financial institutions, wire transfers, credit cards, checks, bank
drafts, the internet and related technologies are meant to keep money-equivalents
sloshing around. 

When central banks create liquidity or pump money into the financial
system, it is like rainfall or snowmelt that feeds rivers and streams
which carry the water around. The centralized institutions , or monetary
policy makers, then become our rain gods or water gods. We do not
know exactly when and how much rain we are likely to receive. But
we have some decent expectations, which is what we call our seasons,
and we have views on what to anticipate based on previous experiences.
Central bank meetings, which do have a fair bit of regularity, to
decide future monetary measures are like the seasonal patterns we
have come to rely on. 

Inflation, which happens when there is too much money in the system,
is like a flood scenario. Drought then becomes a recessionary episode.
Clearly both are not desirable. These are unintended consequences,
both in an economy and other aspects of life, due to the nature of
uncertainty around us, all of which we will discuss in later articles
as it pertains to investment management (Kashyap 2016b). 

Interest rates can be viewed as the ways in which money is taken away
by the system that is designed to send it around. As water flows around,
part of it is lost due to evaporation, seeping into the ground or
flowing into the ocean. This rate at which water is lost by the system
is similar to the base interest rate set by the monetary authorities.
All rates (interest and water loss) and transaction costs are then
modifications of these fundamentals rates specific to different situations.
As our analogy illustrates, when interest rates are higher inflation
will tend to be lower and vice versa. 

There are two main issues with the central monetary or water system.
One boils down to the essence of centralization and the overt dependence
on the main source of water or money, which relies heavily on what
the gods or policy makers do. Central bankers have sole control over
money machines, which have become crucial for financial well-being.
The other issue is that when fresh water does find its way into the
system, the people that can collect most of it are the ones that are
already connected to and well established in the existing network.
In a water system, this is simply the life around rivers and streams
that benefit the most from rainfall. Similarly new wealth ends up
getting concentrated in the hands of those already well entrenched
into the current money transfer mechanisms. 

Unfortunately, the money gods are also likely to be influenced (aka
lobbying) by those that benefit the most whenever new money is printed.
In some ways, the vegetation around a water network precipitates further
rainfall. It would not be entirely incorrect to state that most, if
not all, policy makers have good intentions with no desire to cause
monetary mutilation. We want to emphasize that there are no good or
bad people. Policy makers do what they do, in response to seemingly
tough situations, based on the application of what they have learnt
from mediocre role models. Things have gone haywire due to the lack
of better solutions. The reasons for the lack of superior methods
is due to the need to be conservative when tinkering with economy
wide policies as discussed in Section (\ref{sec:Decrypting-Crypto-and}). 

It is also worth mentioning that the natural system of rainfall or
snowfall and the corresponding watering network, which we should someday
hope to more thoroughly emulate, has no strict parallels for now in
our economy. But the comparison we have outlined serves as a way to
illustrate how the existing monetary system works and to make a strong
case for the necessity of the DeFi technological innovations, which
we discuss next in Sections (\ref{sec:Since-Bitcoin-Was-Coined};
\ref{sec:Decrypting-Crypto-and}). 

\textbf{\textit{Money to Business is as Water to Life. }}

\section{\label{sec:Since-Bitcoin-Was-Coined}Since Bitcoin Was Coined ...}

The invention of Bitcoin in 2008, and the subsequent launch of the
currency in 2009, is no doubt a landmark event permanently etched
in the history of technological innovations. This seminal event is
opening frontiers that are set to transform all aspects of human interactions
(Nakamoto 2008; Narayanan \& Clark 2017; Chen 2018; Monrat, et al.
2019). It has opened the floodgates for innovations seeking to add
different aspects of business and human experiences onto the blockchain. The
rest, as they say, is history.

As the Bitcoin movement gained steam, adding supporters and gaining
momentum as a substitute for money as we know it, many great minds
deemed several improvements as being essential to enhance this landscape.
Ethereum, which was conceived in 2013 and launched in 2015, provided
a remarkable innovation in terms of making blockchain based systems
Turing complete (or theoretically being able to do what any computer
can do: Sipser 2006; End-note \ref{enu:Turing-Complete}). This has
now opened the floodgates for innovators seeking to add different
aspects of business and human experiences onto the blockchain. 

That said, many barriers need to be scaled for the wider adoption
of blockchain technologies. Some of these limitations are: increased
latency, usability limitations, security issues, size and bandwidth
limitations, all of which need to quantified and assessed from a risk
perspective depending on specific use cases (Hughes, et al. 2019).
The funding needs for new projects, and innovators seeking capital
for amazing novel ideas, are creating remarkable opportunities for
investors. Though, the early days of crypto investing are synonymous
with huge swings in prices or volatility, security issues or hacks,
and a lack of protective mechanisms for investors. This has stood
in the way of wider inclusion of crypto currencies, and blockchain
assets, in the portfolios of individuals who do not have a stomach
for roller coaster rides and fatal accidents. 

As the story of Sergey Bubka illustrates (End-note \ref{enu:Sergey-Nazarovich-Bubka}),
human ingenuity has no bounds. Many innovators are learning from the
insightful lessons offered by contemporary chains. Proof of work,
as a consensus mechanism, already has a plethora of interesting alternatives
(Dimitri 2022). Massive efforts are underway to build platforms that
address issues related to high transaction costs, low throughput,
scalability and also to ensure that different chains have a greater
degree of connections and interoperability. 

We see this development of newer chains as a great possibility to
find investment opportunities. Selection of assets will be done across
networks such that each investor can get exposure to the whole suite
of assets on multiple chains. Investing on different chains, and hence
linking different networks, is one way of providing diversified exposure
to an investor base. Though, to trade numerous assets on different
chains can be an onerous task. But, many innovations in decentralized
finance are clearing the way for more investors to enter this space.
Several DeFi protocols are pioneering new methodologies to make crypto
investing less risky, secure and accessible to everyone. Better risk
management techniques will ensure that the gap between funding needs
and the supply of funds will be bridged.

Rigorous risk management within the crypto landscape is something
that is badly needed by all investors seeking crypto exposure. Blockchain
technology is still evolving and this new landscape presents amazing
opportunities to revolutionize all aspects of how we transact. But
several issues, related to security, wild swings in prices and diversification
of assets, have to be addressed for the wider adoption of the blockchain
innovation from an investing perspective. Once investing in this sector
becomes more appealing it will spur further innovation in all other
areas of blockchain technology. 

\section{\label{sec:Decrypting-Crypto-and}Decrypting Crypto and DeFi Investing }

The DeFi phenomenon is offering a radically different paradigm (Werner,
et al. 2021; Xu et al. , 2021). The DeFi movement is creating entirely
new sources and systems of money transfer. This is like tapping into
new and alternate sources of water and building novel techniques to
spread it around. Cryptocurrencies are creating channels that can
stay independent of the centralized systems in many ways. These new
pathways are more accessible for anyone to benefit from them. This
plethora of wealth generation opportunities are due to the many alternate
ways to create and raise money. Technology and other innovations are
also ensuring that these new sources of money have safety measures
designed to prevent inflationary scenarios and several forms of fraudulent
activity. 

This does not imply that sailing on crypto waves with be completely
smooth. There are likely to be unintended consequences in DeFi, just
as in every aspect of life. But a strong argument can be made that
many independently controlled systems are likely to weather tougher
storms, which makes for a more robust overall framework for financial
welfare. Misdemeanour on the part of any DeFi assets will send funds
fleeing to other crypto alternatives that are already there or those
that will mushroom up as needs arise. 

Better solutions are obtained when we can have a trial and error approach
(Kashyap 2021). Such a trial and error approach happens naturally
in the DeFi environment when compared to the full economy. The risk
of any crypto blowing up is unlikely to be fatally detrimental to
the entire system. Taking fundamentally different approaches would
be extremely ill advised in a central banking atmosphere. As the crypto
ecosystem grows, innovators will have greater flexibility in trying
new and unproven techniques. Everyone benefits since the lessons learnt,
even from failed projects, can be applied elsewhere. As many blockchain
projects pioneer the way in bringing sophisticated risk mitigation
principles to the DeFi space, innovation will flourish and continue
to happen in an unperturbed manner. 

\section{\label{sec:Back-To-The-Future-Decentralized-Centralized-Back}Back
To The Future: Decentralized to Centralized and Back }

Money started as a decentralized unit, (in the form of animal skins,
salt, shells etc., to facilitate commerce. It then became centralized
(in the form of coins and notes) when monarchs and later governments
took over the task of supplying currency. Now money, which is increasingly
becoming digital, is moving away from the control of any authority
(Davies 2010; Nakamoto 2008; Reiners 2020; Yadav, et al. 2020). 

\textbf{\textit{History does repeat itself.}}

As the acceptance of crypto currency increases, and the majority of
daily business transactions happen in the alternative world of crypto,
the influence of centralized banking systems and the corresponding
policies will wane. Using our water analogy in this case, as alternate
water sources become important, we can see that the system of rainfall
and other water courses have little effect on our lives. It is a scenario
wherein we (all or humanity or at-least the majority) are living very
far from natural water pathways so that floods and droughts in this
system, have little bearing on us. Clearly we are not there yet both
in the water and monetary system. There is no backup for national
currencies right now. With crypto-currencies on the rise, wealth will
get more options to flee to an alternate asset quite easily. 

Central bank digital currencies (CBDCs) or centralized decentralized
currencies, oxymoronic as it sounds, have many pros and cons (Auer,
et al. 2021; Barrdear \& Kumhof 2021; Agur, et al. 2022). Not wanting
to be left out, many national monetary authorities are planning or
actively contemplating such a scheme. CBDCs can offer stable diversification
benefits and act as a safe haven, if they are governed like other
crypto-currencies with strict guidelines on money supply and other
related aspects. But if just becomes a national currency in digital
form, it will not be very different from other conventional currencies.
If participants are required to follow extensive guidelines before
they can participate, money flow patterns can be traced back to the
originators and CDBC will be less anonymous than pure crypto. Some
participants might favour CDBCs because of the extent of traceability
that comes with recording and displaying all transactions in a blockchain.
But for the same reason, many might stay away from it entirely. This
will be an interesting development to watch once CBDCs become a reality
and try to fit in with the rest of the crypto landscape. 

A counter argument to support centralized currencies can be that it
is easy to manage one currency, whereas too many competing currencies
are bound to cause chaos. To see that this argument holds little merit,
it is important to realize that each crypto-currency is self governing
with members having a transparent view of the policies and in many
cases even having a say on how things should be run. Despite these
measures, things can go wrong sporadically but money will be able
to flee to other sources in such instances given the plenty of alternatives
available. 

There will be turbulence when new and heavy flows of water, or money,
start pouring in. We experience this as volatility in prices when
money moves in and out of crypto-assets. The next wave of innovations
in DeFi will be geared around reducing fluctuations and ensuring the
adoption of crypto as a main stream asset class. 

Numerous startups are pioneering the way by bringing many well established
techniques that have worked well in traditional investing, including
many innovations tailored for the DeFi arena, to decentralized finance.
We list below seven techniques that are essential for Decentralized
Finance and blockchain investing to be more widely adopted by individuals
and businesses alike (Section \ref{sec:Bringing-Risk-Parity}).

\section{\label{sec:Crypto-Conundrums-versus}Crypto Conundrums versus MMT
Mayhem }

Crypto is starting to be perceived as a hedge against a devaluing
dollar (Shahzad, et al. 2020; Blau, Griffith \& Whitby 2021; Conlon,
Corbet \& McGee 2021; Choi \& Shin 2022). Several asset management
firms are actively investing in cryptocurrencies and crypto is being
deemed as an asset class (Hong 2017; Bianchi 2020; Bianchi \& Babiak
2022). The majority of crypto assets are engineered to be actively
deflationary with features such as fixed supply/flow, token burns,
etc. Whether it is going to be completely effective or not is still
to be seen. But the high level of volatility in Crypto assets, and
the general correlation between Bitcoin and the S\&P500, suggests
that a strategy of simply holding crypto assets is not necessarily
a wise move (Chuen, Guo \& Wang 2017; Kosc, Sakowski \& Ślepaczuk
2019; Flori 2019; Xi, O’Brien \& Irannezhad 2020; Liu, Tsyvinski \&
Wu 2022; Troster, et. al. 2019; Figure DELETION:Add figure here).
Nonetheless, if inflation continues to rise then demand for crypto
is likely to go up, driven by corporations wanting to diversify their
reserves. 

There will be another side effect of MMT policies too. As unemployment
falls, there will be more money in people’s pockets and their ability
to save will increase. Some will be attracted by crypto’s get rich
quick headlines, some by the stories of inflation protection. A few
will be drawn by the transparency of DeFi, or in other words, driven
away from banks by the centralized and politicized feel of MMT. 

\subsection{DeFi Yield Farming: The Fields of Gold}

In 2020, many people in the crypto space discovered yield farming:
the ability to increase returns on holdings through different combinations
of staking, liquidity pools, lending, and so on (Xu \& Feng 2022;
End-note \ref{enu:Types-Yield-Enhancement-Services}). Annualized
returns of over 100\%, for a short time anyways,were not uncommon.
It is tempting then, to consider if yield farming will protect against
market fluctuations and environmental shocks. 

In 2020 and 2021, yield farming has achieved higher real yields than
can be achieved by cash or bonds, but all assets don’t behave equally.
Yield farming strategies should be considered as growth assets, highly
dependent upon crypto market volatility and volumes. Crypto deposited
in liquidity pools, such as Uniswap (Angeris, et. al. 2019), earn
a fixed 0.03\% of all trades pro-rata. But the total return depends
on the volume of trades going through the exchange and the capital
is also at risk of impermanent loss (Aigner \& Dhaliwal 2021). High
price volatility, and other fees in some instances, also reduce the
usefulness of yield farming as an inflation hedge. Lending crypto,
on the other hand, is not so volatile. Stable coin (Ante, Fiedler
\& Strehle 2021; Hoang \& Baur 2021; Lyons \& Viswanath-Natraj 2023;
End-note \ref{enu:Stablecoin}) deposits earned yields higher than
bonds or cash in 2021. Rates will vary but they are likely to beat
income from traditional cash deposits under normal circumstances (End-notes
\ref{enu:Lending-rates-on}). Lending crypto can therefore be equated
to the role of bonds in MPT. 

In the year 2021 we found ourselves in an unenviable position. Permanently
low interest rates had broken monetary policy. Quantitative Easing
(QE: Blinder 2010; Fawley \& Neely 2013) had not reached past the
banks, forcing MMT style policies. This massive fiscal stimulus could
backfire by causing steep inflation. MMT centralizes more power in
the hands of politicians who distort the spending patterns, adding
further inflation risk. Evidence of rising inflation will drive further
corporate and consumer demand towards crypto and DeFi. However, crypto
and DeFi are not necessarily immune either. We need a new (or proven,
repurposed) strategy that will offer protection against environmental
shocks. 

Yield farming, however, even with lending, is still at risk from MMT
inflation. Most yield farming is denominated in stable-coins, which
are pegged to the US dollar, so any gains will be subjected to the
same devaluation as the dollar. Cryptocurrencies and blockchain projects
are recent innovations with several active frontiers of research (Yli-Huumo,
et. al. 2016; Xu, Chen \& Kou 2019; Gorkhali \& Shrestha 2020). They
have not yet lived through many different business cycles and stressful
episodes. Reliable data on crypto projects only goes to a little more
than a decade and DeFi platforms are much younger still (Zetzsche,
Arner \& Buckley 2020; Schueffel 2021) with no previous exposure to
inflationary periods, so we do not know for sure how crypto markets
will behave as situations change drastically. 

Some DeFi platforms are attempting to use MPT to construct balanced
crypto indexes (Kim, Trimborn \& Härdle 2021; Lucey, et. al. 2022;Naeem,
et. al. 2022). Although many indices are constructed such that they
can perform reasonably well during a market downturn, MPT does not
defend against environmental shocks (Lee, et. al. 2022; Briola, et.
al. 2023) as we have seen in Section (\ref{sec:MMT-and-MPT}). Another
strategy is required to bring this layer of protection from external
impacts and to construct more robust crypto portfolios. Risk Parity
is such a strategy, which we introduce in the next sub-section (\ref{subsec:The-Decentralized-Ark}).

\subsection{\label{subsec:The-Decentralized-Ark}The Decentralized Ark for The
Great Flood of Post Modern Monetary Maladies}

Risk Parity is an extraordinarily successful methodology from traditional
finance pioneered by Ray Dalio at Bridgewater Associates (Chaves,
Hsu \& Shakernia 2011; Clarke, De Silva \& Thorley 2013). It is specifically
designed to resist environmental factors such as unexpected inflation
and growth (Asness, Frazzini \& Pedersen 2012; Fabozzi, Simonian \&
Fabozzi 2021). 

The traditional finance world has generated many models and innovations
related to trading and risk management. These techniques have gone
several phases of iterations involving implementation and active usage,
which have resulted in many robust and improved techniques becoming
a part of our lives. The challenge is to find ways to simplify many
aspects of the sophisticated techniques used by investment firms and
tailor them to the blockchain environment. 

A big part of our lives revolves around seeking financial security.
The existing mainstream financial industry has done a lot to bring
about financial well-being to many. But there are several issues with
the existing set up. One of the main concerns is that access to financial
services that work really well are highly restricted and not available
to most people. Clearly, we are simplifying the picture significantly
for the sake of this discussion. The essence of what is needed is
about creating ``Equal Wealth Generation Opportunities For Everyone,''
which can be accomplished using decentralized technological innovations
discussed next (Section \ref{sec:Bringing-Risk-Parity}).

To bring effective risk management, and to incorporate asset management
technique such as Risk Parity, to cross-chain DeFi using crypto native
assets, would be to achieve what traditional wealth managers are doing
with stocks and bonds. One approach would be to engineer a set of
four indexes or funds: Alpha, Beta, Gamma and Parity (ABGP). Alpha,
Beta and Gamma are funds with different levels of risk and expected
returns. The investment mandates for these three funds will be to
ensure that, under most circumstances, Alpha will be more risky than
Beta and Beta will be more risky than Gamma. Investors will be able
to combine the three funds depending on their risk appetites. Mixing
Alpha, Beta and Gamma will give the Risk Parity portfolio (Kashyap
2021-X). Risk Parity will be the investment vehicle that will provide
diversified returns, tailored to the risk appetites of each investor,
entirely on a highly secure blockchain environment. Together, ABGP
will capture the market highs, track consistent growth, even out downturns
and protect against shocks. By assigning different weights to each
one, it will provide the capability to offer balanced, risk-adjusted
portfolios. 

\section{\label{sec:Bringing-Risk-Parity}Bringing Risk Parity To The DeFi
Party }

\subsection{A Complete Solution To The Crypto Asset Management Conundrum}

We will publish seven separate articles, that will discuss several
innovations necessary to address the main concerns and to alleviate
the challenges in crypto asset management, in significant detail.
These separate articles, which are referenced in the appropriate place
throughout the text below, contain mathematical formulations and technical
implementation pointers where applicable. The sequence of the next
seven sections will summarize numerous conceptual ideas, in an incremental
fashion, to make DeFi investing more secure and less risky. They will
provide a description of the main components that would need to be
created to reach our goal of bringing Risk Parity to the decentralized
finance world. These articles, which are summarized in separate sections
below, describe our approach to: 
\begin{enumerate}
\item DeFi Security (Section \ref{sec:DeFi-Security:-Turning}; Kashyap
2021-I)
\item Rebalancing and Trade Execution (Section \ref{sec:Trade-Execution:-To};
Kashyap 2021-II)
\item Weight Calculation (Section \ref{sec:VVV-Weight-Calculations:}; Kashyap
2021-III)
\item Risk Parity Construction (Section \ref{sec:The-Risk-Parity}; Kashyap
2021-IV)
\item Profit Sharing and Investor Protection (Section \ref{sec:Caring-for-the-Community};
Kashyap 2021-V)
\item Concentration Risk Indicator and Portfolio Performance Metrics (Section
\ref{sec:Raising-the-Bar-CRI}; Kashyap 2021-VI)
\item Multi-chain expansion and Select Strategic Initiatives (Section \ref{sec:Multichain-Expansion-Select-Strategic};
Kashyap 2021-VII)
\end{enumerate}
\textbf{\textit{Risk Parity will bring long-term stability to DeFi
and the seven innovations we describe below will bring Risk Parity
to the Crypto Party.}}

\subsection{\label{sec:DeFi-Security:-Turning}DeFi Security: Turning the weakest
link into the strongest attraction}
\begin{itemize}
\item We start with the first section of this series of seven, which will
focus on what we consider to be the foremost priority for all organizations
engaged in decentralized finance endeavors, to provide an overview
of a strengthened security blueprint. Kashyap (2021-I) has a detailed
discussion including the corresponding mathematical formulations and
pointers for technological implementation.This first section will
focus on security and the corresponding innovation, which we are calling
the Safe-house. The Safe-house is a piece of engineering sophistication
that utilises existing blockchain principles to bring about greater
security when customer assets are moved around. The Safe-house is
badly needed since there are many ongoing hacks and security concerns
in the DeFi space right now. 
\item Any tall tower, has to withstand a lot of wind resistance. The taller
a structure the stronger the wind forces that it has to overcome.
Hence, the height of the tall tower becomes its weakest aspect. But
if this weakness is addressed properly, and enough safety mechanisms
are incorporated in the design, the height of the tower becomes its
greatest attraction. People flock to the top to marvel at the views
and to admire the accomplishment of having created such a safe and
tall structure. Clearly, the importance of having a solid foundation
for a tall structure cannot be overlooked.
\item Likewise, security is the biggest threat, or the weakest link, in
DeFi right now. DeFi is nothing but the movement of funds seeking
profits. The more the funds move, the greater the security vulnerability.
But if the security concerns are adequately addressed, and appropriate
features are designed to make DeFi investing more safe, this very
weakness can be turned into the greatest attraction. The captivating
fascination will then be the generation of significant wealth for
all participants. The solid foundation, in our case, is the rigorous
risk management, or Risk Parity, that is an intrinsic part of the
framework.
\end{itemize}
DeFi Protocols need to add a protective shield against internal theft
and external intrusion. These are proprietary innovations, entirely
custom built to safeguard our workflow, and we call this, The Safe
House. The Safe House is the combination of a novel software engineering
architecture and automated / manual processes, specific to handling
fund movements, with certain multi-signatory approvals required for
changing key governance policies. This approach will limit any potential
one-time loss to a negligible amount and keep a detailed history of
all the transactions linked to specific internal staff responsible
for fund movements and trade execution. Needless to say, an extra
layer of protection can be provided if the personnel involved in the
process are fully KYC’ed (Know Your Customer or Client; End-note \ref{enu:KYC}). 

\textbf{\textit{Necessity is the mother of all creation / invention
/ innovation, but the often forgotten father is frustration.}}

The enhanced security features we are describe here are, no doubt,
very necessary. But the essence of the security innovations we are
creating are borne out of the numerous troubles several (all?) protocols
are encountering due to unauthorized parties trying to access their
funds (Grobys 2021; Li, et al. 2020). The same could be said about
the rest of the investment vehicles discussed here. These innovations
are very necessary. But the key motivation for these mechanisms and
architectural designs are due to the main issues that one encounters
while trying to obtain: 1) unencumbered access to decent investment
opportunities in the traditional financial world, and 2) peace of
mind while investing in crypto assets.

In today’s blockchain environment, many protocols are constantly under
threat wherein their assets can be taken out or withdrawn by unlicensed
individuals. Cryptographic methods used in blockchain protocols, do
provide a certain amount of security. But, most projects are still
vulnerable either when cryptographic keys, corresponding to fund movements,
are compromised or when internal parties, who have access to the keys,
have the intention of misappropriating investor funds.

The extent of the perils are magnified in the blockchain environment,
since a few parties with malicious intent can reach numerous victims,
given the distributed nature of this technology. This adds to the
perception that security dangers are commonplace and that hackers
are ruling the roost. The many security related incidents stand in
the way of the mass adoption of blockchain technology, which otherwise
has the potential to transform all human interactions. We wish to
do our part to grow this ecosystem by mitigating the harmful influences
and restoring the balance of power to groups that are actively trying
to develop this landscape.

To counter these hazards, we are introducing several new innovations
that will increase the overall defense mechanisms of our protocol.
The novel security innovations, which we are developing, are to ensure
that our system cannot be compromised by either internal or external
actors. Our multi-pronged protection scheme refines the existing cryptographic
cover by adding extra layers of protective shields. By making these
upgrades, we are converting one of the major drawbacks of the DeFi
space to one of the major strengths of our protocol.

The central element of our security innovations is the creation of
a safe house, which will be guarded by private-public key cryptographic
methods, to store all our assets. As an additional measure to enhance
the security, access to the safe house will be provided only upon
verification of the identity of the person requesting the permission.
Our identity verification methodology is above and beyond the security
provided by existing blockchain public-private key cryptographic methods.We
can this technique the One Time Next Time Password (OTNTP). The OTNTP
works similar to the One Time Password (OTP) mechanism. The OTNTP
concept will be used to verify the identity of the portfolio manager,
trying to take out funds, and to allow safe-house access for making
withdrawals. This modified scheme should help with password protection
in decentralized environments where all transaction information has
to be made public for verification purposes.

The safe house has also been designed to detect and neutralize dangers
such as attempts to withdraw by players without the right credentials.
If a real threat is determined, the safe house will go into a locked
state. It will not allow anyone to take out any assets or funds from
it until the severity of the danger has been assessed and it is deemed
safe to resume further operations.

In the event of an extreme situation, such as a malicious party breaching
the safe house, the extent of damage will be limited due to numerous
safeguards on the mobility of funds. This scenario can occur if an
internal member, or an employee, decides to turn rogue. In such a
case, the identity of the person who stole the funds would be established
with certainty, due to our identity verification methodology, and
the amount lost would be minimal. Even if the missing amount is very
small, further action will be taken to recover the lost funds since
the identity of the individual, who took the funds, will be known.

While building the new security features mentioned above, the overriding
challenge will be to ensure that the improved safety procedures will
not become too cumbersome. The objective is to be able to accommodate
more security guidelines and yet operate quickly and effectively to
take advantage of market conditions. This will be discussed further
in the next article, where we consider our trade execution related
innovations. But to summarize, this can be accomplished by matching
fund flows, which are governed by security parameters, to asset management
principles and requirements. The result is a system that will protect
investor assets and yet allow smooth functioning of our investment
machinery.

\subsection{\label{sec:Trade-Execution:-To}Trade Execution: To Trade or Not
To Trade}
\begin{itemize}
\item The portfolio rebalancing mechanism we recommend is based on an innovative
and proprietary system called, The Cascading Waterfall Round Robin
Mechanism. This algorithmic approach recommends an ideal trade size
for each asset during the periodic rebalancing process, factoring
in the gas fee and slippage. 
\item In the hyper-volatile crypto market, our approach to daily rebalancing
will benefit from volatility. Price movements will cause our algorithm
to buy assets that drop in prices and sell as they soar. In fact,
the buying and selling happen only when certain boundaries are crossed
in order to weed out any market noise and ensure sound trade execution.
\item Careful orchestration among mathematical optimization for portfolio
construction, trade automation of the investment apparatus, and human
oversight will allow one to watch out for exceptional situations and
ultimately lead to a better outcome. 
\end{itemize}

\subsubsection{Shakespeare As A Crypto Trader}

\noindent \textbf{\textit{To Trade Or Not To Trade, that is the Question,}}

\noindent \textbf{\textit{Whether an Optimizer can Yield the Answer,}}

\noindent \textbf{\textit{Against the Spikes and Crashes of Markets
Gone Wild,}}

\noindent \textbf{\textit{To Quench One’s Thirst before Liquidity
Runs Dry,}}

\noindent \textbf{\textit{Or Wait till the Tide of Momentum turns
Mild.}}

This is inspired by Prince Hamlet's soliloquy in the works of Shakespeare:
\textquotedbl To be or not to be; that is the question\textquotedbl{}
(End-note \ref{EN:To-Be-Not-To-Be}; Bradley 1991).

We continue with the second of the 7-section series of blockchain
innovations, describing the main components that need to be built,
to get closer to Risk Parity. . In this article, we will take a closer
look at the trade execution innovations we have brought to the DeFi
space in order to rebalance portfolios on a daily basis or even at
an intraday frequency.

“Cascading Waterfall Round Robin Mechanism” are the words we use to
summarize our rebalancing algorithm. To describe how it works, we
first assign a certain capacity to hold funds to each asset in our
portfolio. This capacity is the result of several calculations that
depend upon: 
\begin{enumerate}
\item The risk and return properties of each asset.
\item How the asset prices vary in comparison to other assets in the portfolio.
\item The amount of funds collected for investment (or the total requests
for redemption). 
\end{enumerate}
Once the capacity is determined, we check how much of that capacity
is utilized. This gives us an idea of how much money we can put into
each individual asset when we invest money across our assets. Likewise,
it also tells us how much to pull out of each asset if a withdrawal
is needed. Next, we distribute funds across the assets, or redeem
funds from the assets, in a circular manner, or round robin fashion,
till the full capacity of each asset is reached. As the capacity on
one asset reaches its full limit, the funds start trickling down to
the next asset, similar to a waterfall. The reverse happens when redemptions
are to be fulfilled. Hence the name, “Cascading Waterfall Round Robin
Mechanism”.

After the trade execution schedule is decided, we must consider the
transaction costs of completing the trade orders. There are two main
implicit costs at this stage. First, there are gas fees for each transaction
we execute. Second, there is slippage or market impact. The gas fees
depend on a number of factors, such as the time of execution and the
network on which a trade happens (Zarir, et. al. 2021; Donmez \& Karaivanov
2022). The slippage depends on the size of our trades relative to
the sources of liquidity (Kashyap 2020). Here is a quick summary:
\begin{enumerate}
\item The larger the number of trades, the greater the total gas costs. 
\item The larger the trade sizes, the greater the slippage. 
\item The smaller the trading volume (or liquidity) at the exchange, the
greater the slippage. 
\end{enumerate}
The quintessential trading conundrum in traditional finance is timing
(when to enter a trade) and trade size. The problem is compounded
in crypto since we must factor in the gas fees, which are constantly
variable based on network congestion and type of blockchain. We will
discuss market timing in a later article as it’s a topic particularly
insightful to future front-runners but generally our portfolio will
rebalance daily. The trade size is then determined by the dual objectives
of minimizing both gas fees and slippage. We perform asset level calculations
which are coupled with our “Cascading Waterfall Round Robin Mechanism”
to arrive at recommended minimum and maximum trade sizes. Basically,
the algorithm described in Kashyap (2021-II) will generate (recommends)
a set of min-max values for each trade. 

These trade size recommendations ensure that the fund managers adhere
to the security guidelines, when funds need to be moved into and out
of assets from our secure safehouse. The first section has a detailed
discussion of the security plan (Section \ref{sec:DeFi-Security:-Turning}).
The goal of strengthening security is achieved without creating bottlenecks
for trading since fund movements correspond to trade size restrictions.

The calculation of asset capacities and the rebalancing methodology
are among the most central elements of any investment process. It
is no different in our case. If anything, it is more important for
blockchain projects given the need to adhere to strict risk metrics
and having to incorporate several new techniques geared towards overcoming
the additional challenges in the decentralized space. These two components,
asset capacities and the rebalancing methodology, can be invoked and
utilized on an on-demand basis in the initial stages. The next set
of enhancements are to be able to connect them to data updates, and
completely automate them, so that these calculations can run on a
daily basis or even several times during a 24-hour period.

To govern a system with many moving parts, such as blockchain wealth
management, several parameters must be monitored and tweaked on a
regular basis. The portfolio management team will have to observe
these parameters continuously and update them, as necessary, using
specialized internal tools. The bulk of the configurations that decide
how the system will run are related to asset capacities and trade
executions. In addition, trade executions can be error prone wherein
failures need to be monitored and intelligent customizations to retry
need to be incorporated into the process. Hence trade execution related
parameters and operational procedures will garner significant focus
and a big chunk of time from the investment team. 

The internal tools, to run this operation, are designed such that
the flow of funds happens automatically, for the most part, with human
intervention to complement the decision making. Significant automation
of our investment apparatus will allow us to take advantage of market
opportunities seamlessly and human oversight will enable us to watch
out for exceptional situations and fine-tune the decisions. This coupling
of “Man-and-Machine” will lead to a better final outcome for all our
participants. 

An illustration of this pairing is that our approach to investing
will benefit from volatility, which is seen as the bane of crypto
markets by most players. Volatility, which is the up and down movement
of asset prices, will cause our rebalancing algorithm to buy assets
that drop in prices and sell assets as they start soaring again. But
to filter out the noise, and react only to real signals, the buying
and selling happens only when certain boundaries or range thresholds
are crossed. This spectrum over which transactions happen are automatically
calculated based on asset properties, but fine tuned by investment
specialists. Suffice it to say, while mathematical optimization techniques
offer powerful venues to garner profits, they might fall short of
conquering the extreme scenarios that markets present. Hence mixing
mathematical models with human intuition, that takes care of exceptional
cases, is the ideal recipe for wealth creation.

In the next section (\ref{sec:VVV-Weight-Calculations:}), the third
one, we will go into greater detail regarding the use of risk and
return characteristics to arrive at the capacity for each asset.

\subsection{\label{sec:VVV-Weight-Calculations:}VVV Weight Calculations: Prepared
for the Downside and Primed for the Upside }

\begin{itemize}
\item Two of the most essential ingredients in determining weights are volatilities
and variances (also covariances) of assets. 
\item In the “Velocity of Volatility and Variance” (or VVV) crash protection
mechanism, we adjust the volatilities and the variances (including
covariances) of assets depending on how fast they are likely to change
during market crashes. 
\item Using VVV weights, portfolios can outperform, in terms of returns,
typical portfolios using more conventional weighing mechanisms by
almost 80\% with a considerably higher sharpe ratio (e.g. VVV: 1.91
vs OTHERS: 1.44; Sharpe 1994). We achieve this by taking slightly
higher risk and based on our belief that volatility is a small price
to pay for the convenience of trading anything from anywhere and anytime,
as long as we are sufficiently equipped to deal with downward movements.
\end{itemize}
This third section, of seven planned ones in this series, will provide
a summary of our asset weight calculations. Our novel portfolio weighting
technique considers certain stylized facts about the financial markets.
We have then tweaked the weight computations to factor in the nuances
of the crypto markets.

Researchers have observed and documented, over several decades, a
few stylized facts about traditional financial markets. The propensity
for markets to suddenly crash is much higher than the probability
of an upward movement of similar magnitude (Hong \& Stein 2003; Veldkamp
2005; Bates 2012). When markets crash the prices of most assets move
in tandem or they fall together (Ang \& Bekaert 2004; Hartmann, Straetmans
\& Vries 2004). That is, during market crashes assets tend to have
higher correlations. This correlated movement of prices is due to
the extensive linkages that have developed between financial markets
over the years (Dungey \& Martin 2007). 

Volatilities tend to be higher during market crashes. We need to take
note of this point about asset prices moving a lot more, and tending
to become more volatile when markets crash, as we build a weight calculation
engine. 

Asset weight calculations are generally driven by many inputs, but
the most essential ingredients are volatilities and variances (also
covariances) of assets. We adjust the volatilities and the variances
(including covariances) of assets depending on how fast they are likely
to change during market crashes. Hence, we term this methodology the
“Velocity of Volatility and Variance” (or VVV) crash protection mechanism. 

Our approach is ideally suited for crypto assets which, even during
normal times, are very volatile and are also heavily correlated (Klein,
Thu \& Walther 2018). Hence, we can expect much higher volatilities
and correlations in the crypto investment landscape during a downturn.
Our protection scheme is tailor-made for beating benchmarks when markets
head downward. As we will demonstrate, our methodology also performs
better than other weighting schemes over an entire bull-and-bear market
cycle with no significant underperformance during upward market trends.

Returns are obtained as a direct result of bearing risk. Hence, an
approach that allocates equal risk across all assets will yield a
more robust weighting scheme. This approach is also known as the risk
parity approach. Here, the weights allocated to an asset are proportional
to their corresponding risk (as opposed to their expected return)
relative to that of the overall portfolio. And, this weighting mechanism
is statistically measured using volatility and also accounts for correlations
between assets in a portfolio. 

Implementing risk parity techniques for defi cryptoassets will require
paying special attention to the nuances of how these markets operate.
Crypto markets are more volatile and highly correlated as seen in
Tables (\ref{fig:VVV-Weights-Comparison}; \ref{fig:Correlation-Matrix})
shown in Section (\ref{subsec:Tables-and-Explanations}) below, compared
to traditional finance. A weighting technique designed to outperform
most benchmarks during a market crash while generating solid returns
under other market conditions should be the recommended approach.
And, VVV is our recommended approach. 

The asset weights are the primary constituents required to calculate
asset capacities, which determine how our rebalancing methodology
would work. Section (\ref{sec:Trade-Execution:-To}) has a discussion
regarding our rebalancing methodology. The VVV Weight calculation
algorithms were among the earliest, if not the earliest, components
that we had worked on and tested using historical data. A simple way
to use the weight calculation engine would be to invoke, calculate
and utilize the weights on an on-demand basis. The next set of enhancements
can be able to connect it to data updates, and completely automate
them, so that these calculations can run on a daily basis or even
several times during a 24-hour period.

The VVV weighting methodology factors in several empirical observations
about financial markets and tailors them to the more volatile and
correlated defi environment. This approach does well under a wide
variety of market conditions and is custom built to outperform benchmarks
during market downturns. Building portfolios in this manner epitomizes
our belief that upward movements will take care of themselves, but
it is the downward movements that require the most preparation. As
we have discussed in (Kashyap 2021-III), volatility is caused by the
actions of traders. It is inevitable when a huge number of traders,
with different perceptions of value, transfer large sums of money.
Volatility is a small price to pay for the convenience of trading
anything from anywhere and anytime, as long as we are sufficiently
equipped to deal with downward movements. VVV is the protection mechanism
sorely needed for the DeFi space.

\subsubsection{\label{subsec:Tables-and-Explanations}Tables and Explanations}

Each of the tables in this section are referenced in the main body
of the article. Below, we provide supplementary descriptions for each
table. The full data sample consists of daily observations over the
previous 365 days going back from October 31, 2021. The portfolio
based on VVV significantly outperforms the portfolios using MVO (mean
variance optimization) and MVO without shorts by 74-81\%. Yes, VVV
Portfolio takes more risk (by 34-36\%). However, for the unit of risk
assumed, the portfolio based on VVV will generate a much higher return
- as demonstrated by the higher Sharpe Ratio. This is one trade-off
worth the risk. 

\begin{figure}[H]
\includegraphics[width=16cm,height=10cm]{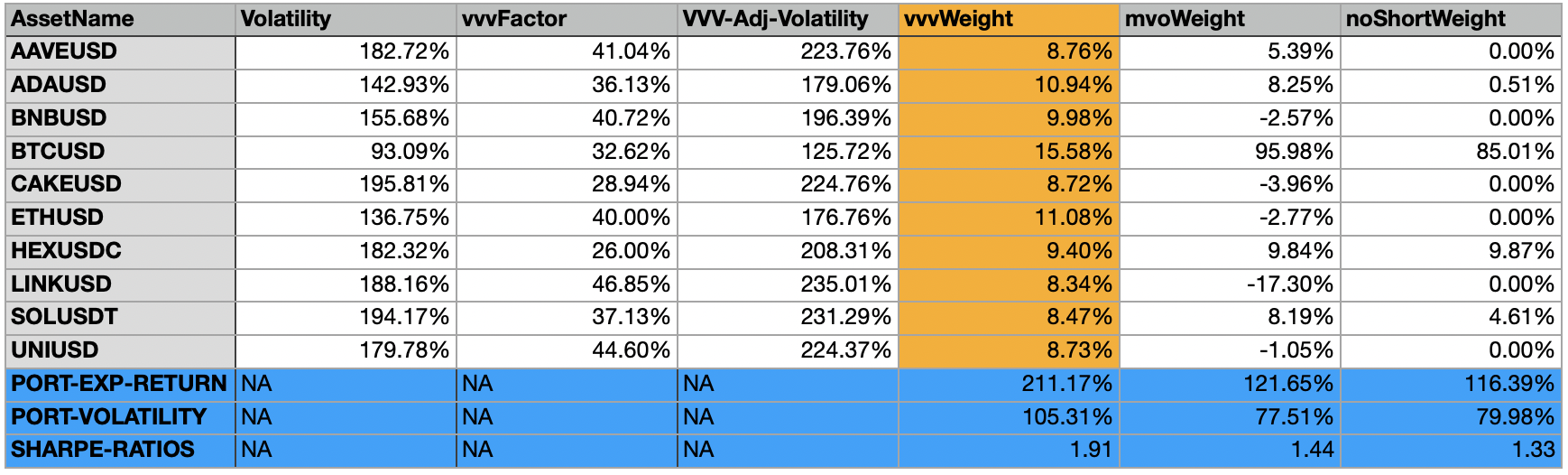}\caption{VVV Weights Comparison to MVO Weights\label{fig:VVV-Weights-Comparison}}
\end{figure}

In the Table in Figure (\ref{fig:VVV-Weights-Comparison}): Each cell
represents the correlation between the asset returns in the corresponding
row and column over the historical period.

In the Table in Figure (\ref{fig:VVV-Weights-Comparison}): The first
column represents the name of the asset under consideration.

The next six columns represent the following information respectively: 
\begin{itemize}
\item \textbf{Volatility} (annualized) of the assets calculated on a 90
day moving internal; 
\item \textbf{vvvFactor} calculated on a 90-day moving interval (i.e. volatility
of volatility); 
\item \textbf{VVV-Adj-Volatility}, which is the sum of annualized Volatility
and vvvFactor; 
\item \textbf{vvvWeight} calculated using VVV-Adj-Volatility; 
\item \textbf{mvoWeight} calculated using mean variance optimization (MVO)
by Markowitz; 
\item \textbf{noShortWeight} calculated using MVO with no shorts (or no
negative weight).
\end{itemize}
The last three rows show the portfolio expected return calculated
based on the annualized average return over the data set horizon;
the portfolio volatility; the sharpe ratio given by the portfolio
expected return minus benchmark rate of 10\% divided by the portfolio
volatility.

\begin{figure}[H]
\includegraphics[width=16cm,height=10cm]{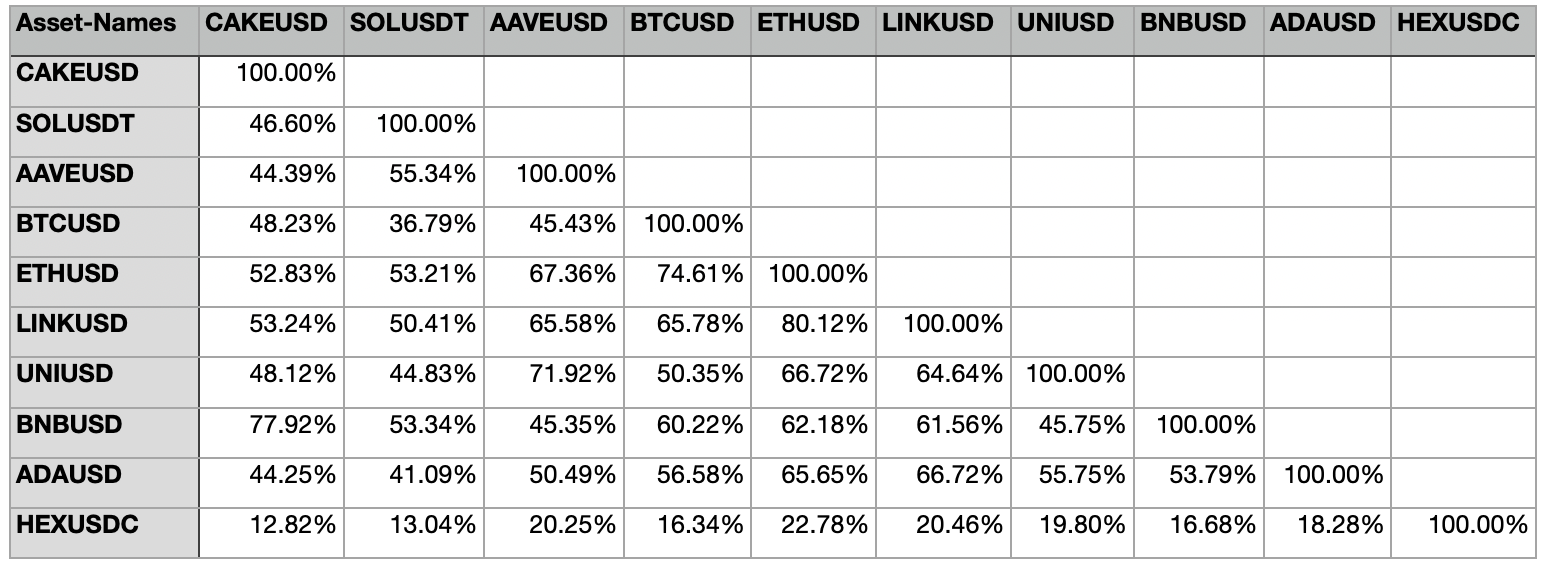}

\caption{Correlation Matrix\label{fig:Correlation-Matrix}}
\end{figure}

\subsection{\label{sec:The-Risk-Parity}The Risk Parity Line: Moving from the
Efficient Frontier to the Final Frontier of Investments}

\begin{itemize}
\item Each of the sub-funds (Alpha, Beta and Gamma, ABG) we discussed in
Section (\ref{subsec:The-Decentralized-Ark}) can be designed to provide
risk parity because the weight of each asset in the corresponding
portfolio can be set to be inversely proportional to the risk derived
from investing in that asset. This can be equivalently stated as equal
risk contributions from each asset towards the overall portfolio risk.
\item Investors can select their desired level of risk or return and allocate
their wealth accordingly among the sub funds (ABG), which balance
one another under different market conditions. This evolution of the
risk parity principle, resulting in a mechanism that is geared to
do well under all market cycles, brings more robust performance and
can be termed as conceptual parity.
\item The inclusion of newer and more diversified assets into the portfolios,
as the crypto landscape expands, can be viewed as a natural progression
from the conventional efficient frontier to a progressive final frontier
of investing, which will continue to transcend itself.
\end{itemize}
Risk Parity is the holy grail that we originally set out to bring
to the decentralized investment world. To obtain parity, the amount
of money allocated to the individual assets in a portfolio has to
be proportional to the extent of risk encountered from investing in
that specific asset. As the risk characteristics of an asset fluctuate,
the weight assigned to that asset has to be correspondingly modified.

A subtle aspect of our portfolio construction and VVV weight calculation
methodology (Section \ref{sec:VVV-Weight-Calculations:}) is that
parity is already accomplished in each of the individual funds Alpha,
Beta and Gamma. These investment products, (Alpha, Beta, Gamma and
Parity) will provide risk managed access to several crypto assets
and strategies. We have adapted many of the well known safety mechanisms
and investor protection schemes that have evolved for several decades
in traditional finance, and combined them with many innovations that
are unique to crypto markets (Section \ref{sec:Caring-for-the-Community}).

Having mentioned that each of the sub funds already achieves risk
parity, we need to draw a distinction between mathematical parity
and conceptual parity. The assets weights are calculated based on
precise rules and mathematical operations and this brings parity to
each of the sub funds at the asset level. While this is still a huge
innovation to bring to the blockchain environment, we wish to proceed
further and bring parity also on a conceptual level.

To elaborate further, we create portfolios that perform satisfactorily
where mathematics can fall short of completely combating market uncertainty.
Broad categories of assets have slightly different risk and return
attributes. By grouping assets with similar responses to different
market regimes, we can ensure that the various groups counterbalance
one another under diverse market conditions. Hence, in addition to
mathematical parity, within each sub fund, each sub fund has an overall
risk return feature which is preferable to the other sub funds under
a particular market criterion.

Another motivation for creating these groups is because even if assets
at the individual level deviate from their risk and expected return
properties, such a misalignment is less likely at the group level.
A few assets in a bunch might display atypical behavior, but the majority
of them will be closer to their representative qualities. The result
is that the overall group can be expected to behave in a certain way
and offset other groups, which are constructed based on the same principle
of clubbing together similar assets, that have different attributes.
We term this fluctuating pseudo-equilibrium between groups of assets
conceptual parity.

A remarkable idea from the financial markets is that of the efficient
frontier (Elton, Gruber \& Padberg 1978; Broadie 1993; Bodnar \& Schmid
2009). There are many ways to combine assets to create portfolios.
Among all the possible combinations the set of combinations that are
superior to the rest, in terms of risk and expected returns, form
the efficient frontier. 

Despite the efficient frontier being an intriguing idea, there are
many practical limitations to accomplish this. To ensure that we are
not constrained by the many reservations, our innovation has been
to come up with the idea of conceptual parity tailored for the crypto
environment (Kashyap 2021-IV). With this modification, Alpha will
be a sub-fund composed of assets that provide higher returns and take
on higher risks. Beta will be representative of the larger market
behavior and provide more steady returns with a correspondingly lower
level of risks. Gamma will take on the role of acting as the risk
free rate, with decent returns but with very little to no risk. Gamma
will also be filled with assets that demonstrate negative correlation
to Alpha and Beta assets. 

The implication of constructing the sub-funds (Alpha, Beta and Gamma)
in this way ensures that when the overall market under performs, which
means Alpha and Beta will not deliver very high returns, Gamma will
still continue to provide acceptable returns because of its negative
correlation to Alpha and Beta. The manufacturing, and linking, of
Alpha, Beta and Gamma will then produce the most efficient set of
portfolios in terms of risk and return characteristics. We term this
collection of portfolios, the parity line.

We believe that the efficient frontier is a moving target, even in
the traditional financial world, with assets being added or removed,
their risk-return properties undergoing alterations and even entire
markets getting transformed. This is all the more the case with the
rapidly evolving crypto landscape, where many new protocols and projects
are appearing on the scene. 

Investment funds will need to add several blockchain protocols, as
they become available, transforming themselves into highly diversified
cross chain collectors of wealth appreciation venues. The plans to
add more protocols will be discussed in the last and seventh section
of this series (\ref{sec:Multichain-Expansion-Select-Strategic}).
Clearly, there will be a need to continuously evaluate new projects,
and if they pass certain due diligence standards, to include them
in the portfolios. Adding exposure to derivative instruments and physical
assets such as gold, real estate, and so on, as and when they become
available. would be prudent as well The implication of this is that
investors will be getting better returns and lower risks, as we seek
out varied sources of risk adjusted returns.

The user experience has to be designed such that investors can tailor
their wealth allocations to their preferred risk appetites. Users
can select either their preferred level of risk or return. Investors
can also directly decide how much of their wealth they want to allocate
to the three funds: Alpha, Beta and Gamma. Once either of the three
routes are selected, (Risk or Return or Weights of Alpha, Beta and
Gamma), the other parameters can be automatically calculated and saved
into an NFT, which the investor will hold for the life of the investment.

The preferences can be changed anytime by investors and this will
trigger a readjustment of their sub fund allocations. Investment specialists
have to also monitor the markets and, as the relationship between
risk and return changes, fine tune the parameters of the parity line
and update the parameters of the portfolio allocations. This will
guarantee that all investors are getting the best possible outcomes
customized for their desired wealth management objectives.

The challenge will be to ensure that the user interactions are intuitive,
and yet their preferences are precisely captured in the investment
decisions. This can be accomplished by letting someone who does not
wish to be bothered with all the settings, or a novice investor, have
the simple option of choosing the default, depositing his funds and
forgetting about everything else. If this is the option chosen, the
portfolio can select a low level of risk and calculate the other parameters
accordingly. Advanced users can choose their risk level or their expected
return, or the weights they want to assign to each of the sub funds.
The other parameters will be automatically calculated.

The outcome of these innovations is an investment machinery that responds
to investor preferences and adapts to changing market conditions.
In addition, these vehicles will adhere to the core tenets of decentralization
and be accessible by anyone. The next, and fifth, section (\ref{sec:Caring-for-the-Community})
will discuss plans to share a significant portion of the profits generated
with the community. All of this can be viewed as a natural progression
from the conventional efficient frontier to a progressive final frontier,
which will continue to transcend itself.

\subsection{\label{sec:Caring-for-the-Community}Sharing is Caring: Setting Aside
Profits for The Crypto Community}

\begin{itemize}
\item A significant portfion of the trading profit will be earmarked for
distribution to the investors who hold project tokens along with investments
in either of the funds: Parity, Alpha, Beta or Gamma. 
\item “Trickle effect” mechanism ensures that the rewards will be paid out
slowly while the amount of the profits designated as performance fees
and the fraction of the fees that will be earmarked for the community
will be varied at different stages of the growth cycle of the project.
\item Each of the four funds (Parity, Alpha, Beta, and Gamma) will operate
as individual profit and loss (P\&L) centers. Investors in Parity
will get their share of the profits which are derived from how their
investment will be split into Alpha, Beta and Gamma.
\end{itemize}
It is essential to stay close to the spirit of decentralization by
setting aside a significant portion (up to 50\%, perhaps) of the trading
profits generated to be paid out to long term investors. This is absolutely
unheard of in the hallowed halls of high finance and a definitive
differentiator in the decentralized community as well. 

The cutting edge designs we have outlined thus far, tailored to overcome
the challenges in the crypto environment, are geared to accumulate
wealth through all cycles in the market. As the investments funds
grow, just like any organization, they will collect fees and generate
revenues to offset the costs. We see two clear claimants to the earnings
produced: the loyal investors and the talented team.

The investors are those who hold the project tokens and also those
who will put capital into Parity, Alpha, Beta and Gamma funds. Without
investors providing capital there is little that could be accomplished
and they need to be rewarded accordingly. Also, the highly skilled
individuals in the project build the vehicles to channel the funds
received into pools that continue to expand. Their efforts are the
key to increase the capital received and due appreciation needs to
be shown. Also, a portion of the proceeds has to be used to develop
the organization and recruit the brightest minds so that it can continue
to do the best towards creating the preeminent wealth management platform
for the masses.

An extremely popular investor protection mechanism in the traditional
finance world is the idea of performance fees and a high water mark
(Goetzmann, et al. 2003; Guasoni \& Obłój 2016; Kashyap 2021-XI).
The simple summary of this concept is that performance fees are charged
only when investors are entitled to a profit off their original principal.
This is perhaps best clarified with a simple numerical illustration. 

For example, let us say an investor deposits 10,000 USD. After some
time, the invested amount grows to 14,000 USD, at which a high water
mark is established. The profit in this case is 4,000 USD. A part
of this profit is taken as performance fees. After this, if the value
of the investment goes down to say 12,000 no performance fees are
charged until the value of investment climbs back above 14,000, the
high water mark. The bottomline is that unless a tangible wealth increase
is generated for every investor, at a holistic level, no performance
fees are paid. This creates a strong incentive for the team to produce
solid returns for the investors.

This simple scenario can get extremely complicated when there are
multiple investors who deposit at different levels of the fund price.
Tracking all this in a smart contract, with the current state of blockchain
technology, is extremely hard and can be deemed almost impossible
(Wang, et. al. 2018; Zou, et. al. 2019; Zheng, et. al. 2020). To be
able to accommodate these complexities we have found a novel solution
that works elegantly, is rather straightforward to implement as a
smart contract, provides the same level of protection to every single
investor and is mathematically identical, in terms of fees and proceeds,
to what investment funds in the traditional world have been doing
for decades. Kashyap (2021-V) has a detailed discussion of this topic
including the corresponding mathematical formulae.

The point worthy of highlighting is that the performance fees, earmarked
for community distribution, will be directed to a separate bucket,
and kept aside, to be paid out regularly to loyal investors. Loyalty
here will be measured in terms of the length of time someone holds
project tokens along with either Parity, Alpha, Beta or Gamma. If
someone has to claim the full share of their reward, they need to
stake project tokens and either Parity, Alpha, Beta or Gamma tokens
and keep it staked to gather rewards. Staking here means deposting
tokens into a smart contract.

The amount of project tokens and other fund tokens to be staked to
claim the full rewards will be dependent on a ratio, such as 1:1 or
2:3 and so on. This is a parameter that can be changed depending on
external factors such as the price of project token, the total investments
made, the amount of profits being generated and so on. To ensure greater
equitability for all investors, who might invest a large sum into
say Parity, if they hold a certain minimum amount of project tokens
they need not adhere to this ratio of project tokens to other tokens
to claim the full share of their profits. The fundamental criteria
is that everyone needs to hold project tokens, in addition to any
other investments they make and they need to continue holding project
tokens, to be eligible to earn their rewards.

Another innovation we have designed, to ensure that investors are
motivated to continue to hold project tokens can be termed the “Trickle
effect”. This mechanism will not pay out the bulk of the profits as
and when they are generated, but the rewards will be paid out slowly.
For example, if the profits we have put into the community pot today
is 100,000 USD. All of this will not be given away on the same day.
Rather, a certain percentage will be paid out today, and a percentage
of what remains will be given out the next day. Let us say, this payout
percentage is 50\%. Then the first day, USD 50,000 will be distributed
to investors as rewards. If no additional profits are generated the
next day, 50\% of what is left will be distributed. So investors will
get 25,000 on the second day and so on. If additional profits are
added to the pool only a percentage of the total accumulated amount
that can be shared will be paid out immediately and the rest will
be retained for subsequent payouts. The result is that there is a
strong incentive to continue to invest in the funds and hold the project
token to be eligible to claim the full stream of profits.

In addition to the above primary utility of the project token, which
enables one to have substantial participation in our upside, there
are two additional reasons for someone to buy and keep the project
token. Holding the project token gives someone the right to participate
in the governance process when the protocol starts to operate as a
DAO, Decentralized Autonomous Organization (End-note \ref{enu:DAO:}).
Also, project tokens owners could be given access to hot, and upcoming
projects with significant upside potential. This would like having
a sub-fund which invests in special projects and accept investments
only from project token owners. Hence the project token will be a
triple utility token.

The amount of the profits designated as performance fees and the fraction
of the fees that will be earmarked for the community can be varied
at different stages of the growth cycle. When significant profits
are being generated, the possibility of using those proceeds to burn
(retire) some of the project tokens (from the circulation) will be
pursued so that it might act as a deflationary mechanism and prop
up the token price. When the platform starts to function as a DAO,
some of these governance parameters will be subject to community input.

The bulk of the revenue generated will be from performance fees. Setting
aside, a big chunk of this for the community might seem excessive.
To toe the fine line between growth (investment for future) and decentralization
(distribution of profit to the community for now), during the initial
stages of the lifecycle before transforming into a DAO, the parameters
can be skewed towards favoring growth. And even at a later stage,
there can be a threshold in the distribution bucket, so that profits
only in excess of that level will be handed out using the trickle
effect. This threshold can be varied depending on the magnitude of
profits, the stage of growth and future plans, risk provisions to
accommodate unforeseen emergency funding requirements, and to ensure
that both investors and employees are compensated fairly for their
contributions.

Since Parity is a combination of Alpha, Beta and Gamma. Investors
in Parity will get their share of the profits which are derived from
how their investment will be split into Alpha, Beta and Gamma. Each
of the four funds will continue to operate as individual profit and
loss (P\&L) centers. Since claiming profits requires holding both
project tokens and one of the other tokens, the total rewards will
match and exceed the rewards from holding project tokens alone. In
the initial phases when very little profits are being generated investors
can deposit project tokens in a single sided staking pool, which can
eventually be phased out entirely and replaced by the enhanced profit
sharing plan discussed above.

This profit sharing mechanism, and related innovations, will ensure
that we are creating a strong economic incentive for investing in
and holding fund tokens (Alpha, Beta, Gamma, and Parity), and especially
the project governance token. The other unintended, yet perhaps welcome
consequence, will be that such an approach might become the trendsetter
clearing the path for other enterprises to be able to share their
proceeds with all their stakeholders, which is the true hallmark of
decentralization.

\subsection{\label{sec:Raising-the-Bar-CRI}Raising the Bar for Portfolio Performance
Measurement: The Concentration Risk Indicator}

\begin{itemize}
\item CRI (the concentration risk indicator), modified and adapted from
the Herfindahl-Hirschman (HH) index, is a novel risk measurement measure
we have developed. Supplementing the CRI with other metrics allows
us to gauge how portfolios are performing and to compare them to the
wider set of crypto investment opportunities.
\item Asset selection guidelines, due diligence process, risk management
oversight and the VVV weighting methodology (Section \ref{sec:VVV-Weight-Calculations:})
take care of monitoring the many other factors that dictate whether
an asset makes a good investment.
\item Continuous innovation, inspired by how world class athletes deal with
new record settings, is the hallmark of any outstanding investment
management approach. 
\end{itemize}
Bringing Risk Parity to the DeFi Party has been the impetus for numerous
innovations and designs described here. Once these novel techniques,
are implemented in the blockchain environment, it will create an unparalleled
platform for wealth generation accessible by anyone. 

In this sixth section, we will discuss a new metric we have developed,
termed the concentration risk indicator (CRI), that will allow us
to gauge how portfolios are performing and compare them to the wider
set of crypto investment opportunities. This metric is focused on
the current facet of the decentralized terrain, wherein the majority
of the wealth is restricted to a small number of tokens. Our new measure,
when supplemented with other well known portfolio measurement yardsticks,
will give a complete picture of how well any investment machinery
is working.

The crypto landscape has many individuals who invested early in projects
such as Bitcoin and Ethereum, when they were up and coming prospects.
These holdings have grown significantly to become fairly large positions.
From a portfolio perspective, their wealth is heavily concentrated
in a few names. This is also the very nature of the crypto markets,
where bitcoin and ethereum command more than 60\% of the total market
capitalization.

The number of tokens listed now on major data providers, such as coinmarketcap,
is around 19,500+ as of May-25-2022 (End-note \ref{enu:CoinMarketCap}).
This figure has more than doubled within the last one year. With a
trend where several tokens appear, and an equal or greater number
disappear, choosing the right investments is an arduous task. Proper
due diligence and research procedures need to be utilized for forming
portfolios.

Having the right selection methodology is crucial and, once a selection
is made, evaluating the corresponding performance is equally important.
To address drawbacks with prevailing methodologies, and to supplement
existing methods, we had to come up with the CRI.

The concentration risk indicator is meant to indicate how diversified
the holdings in a portfolio are. This is a modification of the Herfindahl–Hirschman
(HH) index, (Rhoades 1993; End-note \ref{enu:The-Herfindahl-index}),
which is widely used as a measure of the size of firms in relation
to the industry they are in and an indicator of the amount of competition
among them. We tailor the HH-Index to the crypto markets based on
the following two features: 1) the larger the market cap of an asset,
the lesser the risk of holding it; 2) the more volatile an asset,
the higher the risk of holding it. The amount of money invested in
an asset as a fraction of the overall wealth held by an investor,
which is also the weight of the asset within the portfolio, is also
factored in this metric.

The concentration risk indicator can be calculated for individual
assets and for portfolios of assets as well. When comparing two investments,
the lower the concentration risk, the better the investment from a
diversification point of view. If two assets have comparable market
cap then the asset with lower price volatility would be preferred.
Instead of using the raw market cap values, we normalize and express
it as a fraction to the total crypto market cap before including this
factor in the concentration risk indicator. Similarly, if two assets
have comparable levels of volatility, the asset with a greater share
of the market would be preferred.

As an illustration, given a choice between holding BTC or ETH, if
we need to isolate the effect of size on our investment, BTC with
its higher market capitalization would seem as a better alternative
(End-note \ref{enu:Crypto-Ranking}). Likewise, SOL, XRP and ADA have
a similar level of market share and hence their price volatility determines
how concentrated an investment in these assets would be. This simplified
example is meant only to illustrate the influence of size and volatility.
Clearly, ETH has many other features that could potentially qualify
it as a more desirable investment than BTC. An argument can also be
made that tokens with higher market cap will have lower volatility
than the ones with lower market cap (Fama \& French 1992; Perez‐Quiros
\& Timmermann 2000; Van Dijk 2011; Fama \& French 2018).

The first draft of this article originally included LUNA along with
SOL, XRP and ADA in the above example. But the events of the past
few weeks are a wake up call to all players in the Crypto landscape.
Better risk management, stress testing and checking numerous seemingly
unlikely scenarios are an absolute necessity. The recent LUNA / UST
episode on the Terra network, from May 8 to May 13 2022 and beyond,
is a demonstration of the risk of holding concentrated portfolios
(Uhlig 2022; Lee, et. al. 2022; Briola, et. al. 2023).

We have been developing and testing this new CRI metric for several
months now. The creation of this measure was to have a numeric score
to show people that no concentrated holding is safe even if it is
as large as BTC or ETH. Clearly the recent events, surrounding LUNA,
have not been easy for many of us. But it affirms our long held belief
that nothing can be taken for granted in crypto, and for that matter
anywhere, and suitable risk mitigation plans have to be made even
for rather extreme scenarios. These beliefs are encoded in the risk
management guidelines, espoused in Kashyap (2022), that investment
teams have to adhere to. Going beyond just this new metric, a rigorous
approach to investing and risk management is what investing on blockchain
needs. Risk parity and the whole suite of tools we are describing
are exactly the need of the hour.

While the CRI metric gives preference for larger and more stable assets,
smaller and newer assets will be the drivers of growth. Hence, adding
a greater number of smaller assets can compensate for the risk they
bring in terms of size. Asset selection guidelines, due diligence
processes, proper risk management oversight and our VVV weighting
methodology take care of monitoring the many other factors that dictate
whether an asset makes a good investment.

We are deeply cognizant of the delicate necessity that to keep improving
the performance of our portfolios, our tools to assess performance
need to keep improving as well. Volatility, returns and other metrics
are generally more meaningful when evaluated on a comparative or relative
basis. Since Crypto investments are deemed riskier, to perform a proper
comparison for risk and return, it will be helpful to try to incorporate
benchmarks external to the crypto world. Initially it would be easier
to start displaying returns, volatility and the concentration risk
indicator, over different time intervals, comparing the main funds
we have discussed, Alpha, Beta and Gamma, with several other prominent
crypto assets. Parity investments at different risk levels can also
be viewed as different crypto funds and similar comparisons can be
performed.

At a later stage, we can compare the volatility of crypto investment
funds to an external benchmark such as the VIX volatility index (End-note
\ref{enu:Cboe-Global-Markets-VIX}; Wang 2019). Also returns can be
benchmarked against returns from other asset classes outside the crypto
landscape. External indices across asset classes such as stocks, bonds,
commodities and so on could be useful for this purpose. There are
further improvements we are planning to the CRI, so that it will take
into account the proportion of assets invested on different chains.
Similar to the basic CRI discussed earlier, the enhanced CRI will
reflect the diversification benefits of amounts invested across multiple
chains (Kashyap 2021-VI).

A big part of crypto investing, and also perhaps many other aspects
of our lives, is dealing with uncertainty and our struggle to overcome
it. Sergey Bubka is our Icon of Uncertainty. As a refresher, he broke
the pole vault world record 35 times (End-note \ref{enu:Sergey-Nazarovich-Bubka}).
Pole vault is a simple sport, where you use a long pole to jump over
another long pole, which is placed on top of two other long poles. 

Applying the central idea from pole vault to the crypto landscape,
we can view the introduction of any new trading strategy or innovation
or even regulatory change as equivalent to the raising of the bar
in the game of pole vault. Once a new innovation starts becoming popular
others imitate it or come up with other wonderful ideas, and we need
to find ways to better ourselves. Each time the bar is raised the
spirit of Sergey Bubka, whom we admire a lot and who is a huge inspiration
for us, will help us to reach higher and find a way over the raised
bar. 

This anecdote, about Sergey Bubka and overcoming uncertainty, forms
our fundamental belief that galvanizes us to constantly innovate and
find better models, metrics, trading strategies and ways to generate
wealth for all investment participants.

\begin{figure}[H]
\includegraphics[width=18cm]{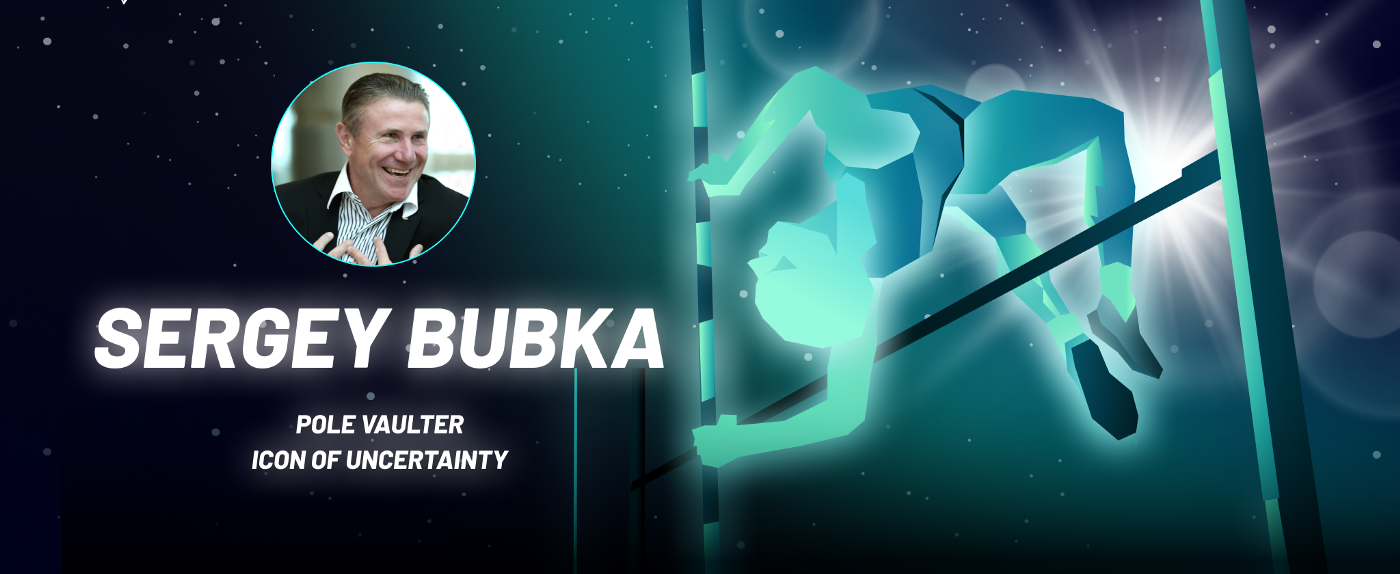}\caption{\label{fig:Sergey-Bubka:-Icon}Sergey Bubka: Icon of Uncertainty}
\end{figure}

\subsection{\label{sec:Multichain-Expansion-Select-Strategic}Multichain Expansion
and Select Strategic Initiatives: Building Bridges That Do Not Burn}
\begin{itemize}
\item Selection of assets can be done across networks in such a way that
each of the investors will get exposure to the whole suite of assets
we have on all chains. Investing on different chains, and hence linking
different networks, is a way of providing diversified exposure to
the investor base. The fund prices can be the same across all the
networks where the investments are deployed. 
\item Use of bridges should be cautious at first and depending on asset
flow requirements, and improvements to the corresponding infrastructure,
we can readjust the fund transfer limits.
\item Our all inclusive approach is to recognize the team and community
as one group: Our Human Capital. Essentially what this means is that
we will not differentiate between the team and community but instead
view them simply as different subunits or divisions within our organization.
\end{itemize}
In this seventh and the final section of the series, we will touch
upon some strategic plans, that projects can consider over the long
term horizon, and our motivations for choosing this particular set
of initiatives. A key focus that will be highlighted are the efforts
and the rationale for rolling out various investment products on different
chains.

As discussed in Section (\ref{sec:Since-Bitcoin-Was-Coined}), many
protocols with wonderful possibilities are being developed. At this
time, ETH, BSC and Polygon are good candidates for an initial launch
of the investment funds, including risk parity components and a safe
house. Lauching the product in phases is practical so that we can
thoroughly test on each platform and resolve any issues related to
each blockchain system. These three protocols are good candidates
for starting out given the remarkable progress they have made, the
stability they bring to this space and the similarity they offer in
terms of technological requirements. All three of them are EVM (Ethereum
Virtual Machine) compatible, making it relatively straightforward
to start using another of these platforms once a product is built
for one of these chains (Jia \& Yin 2022).That said: ETH with high
gas fees, BSC with some vulnerabilities in its choice of validators,
Polygon with scalability issues at times represent challenges that
are inherent in any technology saga. Numerous small tweaks and entire
redesigns of architectural frameworks are being undertaken with these
networks and their future looks promising.

To elaborate on this further, fund prices will be the same across
all the networks on which the investment infrastructure will be deployed.
To arrive at a fund price, we will consider two factors: 1) the combined
total value locked (total investment funds received) across all networks
for that fund, and 2) the number of tokens issued for that fund across
all networks. For example, if an investor invests \$50,000 USD on
only one network, say BSC for example, he / she will be getting exposure
to the performance and diversification benefits of all the assets
held across all networks in that fund. For an investor to do this
by himself would be an extremely arduous task. For any one person
to continually monitor such a portfolio spread across networks, and
change it based on market conditions, would be almost impossible.

Solana, Fantom, Harmony One, Avalanche are some chains, which are
showing a lot of promise, and should feature actively in any plans
to deploy products and invest in assets on these platforms. Several
other platforms could also be on the immediate radar. As and when
promising investment opportunities arise on newer chains, it is prudent
to be prepared to monetize that.

From a network exposure point of view, the entire amount of funds
under management will be seen from two perspectives: one is the network
portfolio and a global portfolio that aggregates all of the network
portfolios. We need to monitor the weights of assets globally and
strict risk management limits have to apply to the global portfolio.
This global capacity on each asset will be filled by positions on
each network depending on how easily funds can flow between networks.
The amount of funds we transfer across networks via bridges will depend
on the capacity of the bridge that spans across networks, the relative
gas fees of the networks, the amount of funds we receive, the asset
availability, and the exposure we assign to each network.

Right now, “Bridges” built between various networks are both a “Bottleneck”
and an “Achilles heel”. They limit the amount of funds that can move
between networks and they also become an attack vector for hackers
to target, resulting in the loss of funds (End-note \ref{enu:Blockchain-Bridges};
Belchior, et. al. 2021; Lee, et. al. 2022; Li, Liu \& Tan 2022). Hence
the use of bridges should be cautious at first and depending on asset
flow requirements, and improvements to the corresponding infrastructure,
we need to readjust our fund transfer limits.

There will be different degrees of correlation between prices across
different chains depending on the extent of inter-connectedness between
them. As the fund flow increases across existing chains, it is highly
likely that the movements will increase in lock steps. The greater
overlap between chains in terms of asset movements will also bring
about the risk for a drastic drop in total value invested in on any
chain, if that particular network starts to lose trust and get abandoned.
Initially, frictions that will impede fund movements will serve the
best interest of certain parties. But as competitive pressures erode
the frictions, they will later exacerbate certain other risks.

Trials with small errors are the key to sustained progress. As product
iterations happen we have to continuously assess what we have learned
so far and look to make improvements. Huge mistakes such as Terra
/ Luna, unintended and unwanted as they might be, can be quite costly
to the system. It is reassuring to see that efforts are being made
towards a recovery and the community of blockchain enthusiasts are
not deterred by this setback. This resilience is to be whole heartedly
applauded. One of the core reasons why we have written this article
is to be prepared for such drastic incidents by incorporating more
robust mechanisms. Despite the unfortunate losses for many victims
it serves to affirm that better risk management principles, and the
other benefits described here, are sorely needed by the blockchain
community. 

Another set of bridges that need to be actively built are strategic
partnerships to ensure that the crypto environment can be highly inclusive,
and connect investing to several real world platforms, solving many
problems that plague humanity along the way. These will be ongoing
and some focus on these initiatives will be required once the main
products are tested thoroughly and deployed.

Moving on to the investing activities and plans for seeking additional
returns. As time goes on several overlay strategies can be added,
to the basic funds (ABG) we have discussed in Section (\ref{subsec:The-Decentralized-Ark}),
and we can seek to benefit from any potential opportunities that open
up (Mulvey, Ural \& Zhang 2007). A team of researchers and investment
specialists need to continually scour the landscape to identify ways
to generate profits. The development of new networks, and derivative
providers within networks, will enable us to use options as a hedging
mechanism (Hull 2003). This will help to protect from market crashes
and can be used to reduce the risk in Gamma. Also, derivative strategies
combined with rigorous risk management can help to gain additional
returns (Huberts 2004; Madan \& Sharaiha 2015). These will be considered
for Alpha.

Further overlays can be based on specific allocations to sectors we
see as promising. This would be similar to sector themed sub-indices
or ETFs but within a larger grouping of assets (Healy \& Lo 2009;
Mohanty, Mohanty \& Ivanof 2021). These developments can be part of
Beta, including allowing investors to customize their preferences
in a basket or theme. Initially it will be easier to accept investments
made only in stable coins (USDT, USDC and BUSD). We are developing
mechanisms through which investors can participate in these investment
vehicles by making deposits denominated in a larger set of assets
(Kashyap 2021-IX).

DAOs (Decentralized Autonomous Organizations: End-note \ref{enu:DAO:};
Kashyap 2021-VIII) are seen as the way forward for blockchain systems.
Tao or Dao is the natural order of the universe (End-note \ref{enu:TAO:}).
This intuitive knowing of life cannot be grasped as a concept. Rather,
it is known through actual living experience of one's everyday being. 

We draw inspiration from the natural order of the universe, as we
know it, and try to incorporate these principles into the DAO, as
wish to create it. The simple message that we arrive at is this: our
engagement with the DAO has to be one of absolute involvement, just
like our experience with the universe has to be. A belief that a DAO
is not just an organization but a way of life is necessary to ensure
that we can make the most of the fascinating capabilities of the decentralized
realm. We propose several mechanisms in Kashyap (2021-VIII) by which
such an attitude can be developed among all the participants and how
technological systems can facilitate that.

The topic of the DAO can be a seven hundred part, or more, series
of articles. But in this article, we summarize the most essential
ingredients for a DAO. The core philosophy we espouse regarding this
topic is that we need to have similar principles, and the same group
of people in the organization, for all human touch points and to handle
the policies for both the Team and the Community. Having such a structure
would be a huge first step towards the establishment of a DAO. 

To the best of our knowledge every organization thus far, both within
the crypto terrain and outside, has viewed employees and clients,
including external stakeholders, as two separate entities. The all
inclusive approach that we recommend is to recognize the team and
community as one group, the human capital. Essentially what this means
is that we should not differentiate between the team and community
but instead view them simply as different subunits or divisions within
the organization. The team and community will include all human (and
perhaps, even non-humans at a later stage) actors that are participating
in some aspect of the DAO. 

It is important to have similar principles, but not the same ones,
for all participants. Though, similar to any other organization, there
will be different sets of responsibilities and rewards for different
departments within the organization. We need to have different incentives,
and guidelines, for the many different duties that members of the
human capital perform. Surely, such a unique approach might bring
conflicts that are inherent when attempting such a radical change.
The objective should be to establish this paradigm as an intrinsic
part of the culture and revise policies to ensure conflicts are minimized.

\section{\label{sec:Crypto-or-Cash}Crypto or Cash or Crypto will become Cash}

We have considered the many challenges in blockchain projects and
decentralized finance. We have discussed several innovations that
will aid DeFi projects in their efforts to become more widely adopted
by the general public. The innovations we have described here are
covered in greater detail, including mathematical formulations and
technical implementation pointers in separate articles, in separate
articles which are referenced at the appropriate places. As with any
technology that holds vast promise, it is hard to accurately pin down
exactly how it will shape our lives. That said, money is likely to
end up almost entirely in a digital format. How soon will the corresponding
developments democratize wealth management is a question that we need
to ponder further upon?

Blockchain technology is creating a fascinating marketplace where
any-one can participate from any-where and at any-time to trade any-instrument
to full-fill almost any-desire. Preparation for the unintended and
unwelcome outcomes, by coupling rigorous risk mitigation with continuous
innovation, will ensure that this technology fulfills the massive
potential it holds. The possibilities are endless.

\section{End-notes }
\begin{enumerate}
\item \label{enu:Decentralized-finance}Decentralized finance (often stylized
as DeFi) offers financial instruments without relying on intermediaries
such as brokerages, exchanges, or banks by using smart contracts on
a blockchain. \href{https://en.wikipedia.org/wiki/Decentralized_finance}{Decentralized Finance (DeFi), Wikipedia Link}
\item \label{enu:Types-Yield-Enhancement-Services}The following are the
four main types of blockchain yield enhancement services. We can also
consider them as the main types of financial products available in
decentralized finance: 
\begin{enumerate}
\item \label{enu:Single-sided-staking-allows}Single-Sided Staking: This
allows users to earn yield by providing liquidity for one type of
asset, in contrast to liquidity provisioning on AMMs, which requires
a pair of assets. \href{https://docs.saucerswap.finance/features/single-sided-staking}{Single Sided Staking,  SuacerSwap Link}
\begin{enumerate}
\item Bancor is an example of a provider who supports single sided staking.
Bancor natively supports Single-Sided Liquidity Provision of tokens
in a liquidity pool. This is one of the main benefits to liquidity
providers that distinguishes Bancor from other DeFi staking protocols.
Typical AMM liquidity pools require a liquidity provider to provide
two assets. Meaning, if you wish to deposit \textquotedbl TKN1\textquotedbl{}
into a pool, you would be forced to sell 50\% of that token and trade
it for \textquotedbl TKN2\textquotedbl . When providing liquidity,
your deposit is composed of both TKN1 and TKN2 in the pool. Bancor
Single-Side Staking changes this and enables liquidity providers to:
Provide only the token they hold (TKN1 from the example above) Collect
liquidity providers fees in TKN1. \href{https://docs.bancor.network/about-bancor-network/faqs/single-side-liquidity}{Single Sided Staking,  Bancor Link}
\end{enumerate}
\item \label{enu:AMM-Liquidity-Pairs}AMM Liquidity Pairs (AMM LP): A constant-function
market maker (CFMM) is a market maker with the property that that
the amount of any asset held in its inventory is completely described
by a well-defined function of the amounts of the other assets in its
inventory (Hanson 2007). \href{https://en.wikipedia.org/wiki/Constant_function_market_maker}{Constant Function Market Maker,  Wikipedia Link}

This is the most common type of market maker liquidity pool. Other
types of market makers are discussed in Mohan (2022). All of them
can be grouped under the category Automated Market Makers. Hence the
name AMM Liquidity Pairs. A more general discussion of AMMs, without
being restricted only to the blockchain environment, is given in (Slamka,
Skiera \& Spann 2012).
\item \label{enu:LP-Token-Staking:}LP Token Staking: LP staking is a valuable
way to incentivize token holders to provide liquidity. When a token
holder provides liquidity as mentioned earlier in Point (\ref{enu:AMM-Liquidity-Pairs})
they receive LP tokens. LP staking allows the liquidity providers
to stake their LP tokens and receive project tokens tokens as rewards.
This mitigates the risk of impermanent loss and compensates for the
loss. \href{https://defactor.com/liquidity-provider-staking-introduction-guide/}{Liquidity Provider Staking,  DeFactor Link}
\begin{enumerate}
\item Note that this is also a type of single sided staking discussed in
Point (\ref{enu:Single-sided-staking-allows}). The key point to remember
is that the LP Tokens can be considered as receipts for the crypto
assets deposits in an AMM LP Point (\ref{enu:AMM-Liquidity-Pairs}).
These LP Token receipts can be further staked to generate additional
yield.
\end{enumerate}
\item \label{enu:Lending:-Crypto-lending}Lending: Crypto lending is the
process of depositing cryptocurrency that is lent out to borrowers
in return for regular interest payments. Payments are typically made
in the form of the cryptocurrency that is deposited and can be compounded
on a daily, weekly, or monthly basis. \href{https://www.investopedia.com/crypto-lending-5443191}{Crypto Lending,  Investopedia Link};
\href{https://defiprime.com/decentralized-lending}{DeFi Lending,  DeFiPrime Link};
\href{https://crypto.com/price/categories/lending}{Top Lending Coins by Market Capitalization,  Crypto.com Link}.
\begin{enumerate}
\item Crypto lending is very common on decentralized finance projects and
also in centralized exchanges. Centralized cryptocurrency exchanges
are online platforms used to buy and sell cryptocurrencies. They are
the most common means that investors use to buy and sell cryptocurrency
holdings. \href{https://www.investopedia.com/tech/what-are-centralized-cryptocurrency-exchanges/}{Centralized Cryptocurrency Exchanges,  Investopedia Link}
\item Lending is a very active area of research both on blockchain and off
chain (traditional finance) as well (Cai 2018; Zeng et al., 2019;
Bartoletti, Chiang \& Lafuente 2021; Gonzalez 2020; Hassija et al.,
2020; Patel et al. , 2020). 
\end{enumerate}
\end{enumerate}
\item \label{enu:United-States-Department-Bond-Yields}United States Department
of Treasury provides daily interest statistics for the past several
decades: \href{https://home.treasury.gov/policy-issues/financing-the-government/interest-rate-statistics}{US Department of Treasury, Interest Rate Satistics}.
\item \label{enu:Money-Machines}Money Machines will get turned off, as
soon as people step in to take advantage of it. This is also know
as arbitrage (Shleifer \& Vishny 1997) and it is possible when the
law of one price is violated (Isard 1977; Crouhy-Veyrac, Crouhy \&
Melitz 1982).
\item \label{enu:Turing-Complete}In computability theory, a system of data-manipulation
rules (such as a computer's instruction set, a programming language,
or a cellular automaton) is said to be Turing-complete or computationally
universal if it can be used to simulate any Turing machine. \href{https://en.wikipedia.org/wiki/Turing_completeness}{Turing Completeness,  Wikipedia Link}

A Turing machine is a mathematical model of computation describing
an abstract machine that manipulates symbols on a strip of tape according
to a table of rules. Despite the model's simplicity, it is capable
of implementing any computer algorithm. \href{https://en.wikipedia.org/wiki/Turing_machine}{Turing Machine,  Wikipedia Link}
\begin{doublespace}
\item \label{enu:Sergey-Nazarovich-Bubka}Any attempt at regulatory change
is best exemplified by the story of Sergey Bubka, the Russian pole
vault jumper, who broke the world record 35 times. Attempts at regulatory
change can be compared to taking the bar higher. Similarly, when faced
with obstacles, or constraints or problems, the spirit of Sergey Bubka
within all of us will find a way to surmount those challenges and
sail over them.
\end{doublespace}
\begin{itemize}
\begin{doublespace}
\item Sergey Nazarovich Bubka (born 4 December 1963) is a Ukrainian former
pole vaulter. He represented the Soviet Union until its dissolution
in 1991. Sergey has also beaten his own record 14 times. He was the
first pole vaulter to clear 6.0 metres and 6.10 metres. Bubka was
twice named Athlete of the Year by Track \& Field News and in 2012
was one of 24 athletes inducted as inaugural members of the International
Association of Athletics Federations Hall of Fame. \href{https://en.wikipedia.org/wiki/Sergey_Bubka}{Sergey Bubka, Wikipedia Link}
\end{doublespace}
\end{itemize}
\item \label{enu:Stablecoin}A Stablecoin is a type of cryptocurrency where
the value of the digital asset is supposed to be pegged to a reference
asset, which is either fiat money, exchange-traded commodities (such
as precious metals or industrial metals), or another cryptocurrency.
\href{https://en.wikipedia.org/wiki/Stablecoin}{Stable Coin,  Wikipedia Link}
\item \label{enu:Lending-rates-on}Lending rates on Stable coins have fallen
down from around 12\% in 2021 to less than 3\% in 2022 after the crash
of LUNA and FTX (Uhlig 2022; Fu, Wang, Yu \& Chen 2022). The stable
coin lending rates mentioned here are from Aave, a DeFi protocol on
the Ethereum network (Ao, Horvath \& Zhang 2022).
\item \label{enu:KYC}The know your customer or know your client (KYC) guidelines
in financial services require that professionals make an effort to
verify the identity, suitability, and risks involved with maintaining
a business relationship. \href{https://en.wikipedia.org/wiki/Know_your_customer}{Know Your Customer,  Wikipedia Link};
\href{https://www.investopedia.com/terms/k/knowyourclient.asp}{Know Your Client,  Investopedia Link}
\item \label{EN:To-Be-Not-To-Be}To be, or not to be\textquotedbl{} is the
opening phrase of a soliloquy given by Prince Hamlet in the so-called
\textquotedbl nunnery scene\textquotedbl{} of William Shakespeare's
play Hamlet, Act 3, Scene 1. (William Shakespeare: \href{https://en.wikipedia.org/wiki/William_Shakespeare}{William Shakespeare, Wikipedia Link})

To be, or not to be, that is the question: 

Whether 'tis nobler in the mind to suffer 

The slings and arrows of outrageous fortune, 

Or to take Arms against a Sea of troubles, 

And by opposing end them: to die, to sleep ...
\item \label{enu:DAO:}A decentralized autonomous organization (DAO) is
an organization constructed by rules encoded as a computer program
that is often transparent, controlled by the organization's members
and not influenced by a central government. \href{https://en.wikipedia.org/wiki/Decentralized_autonomous_organization}{Decentralized Autonomous Organization, Wikipedia Link}
\item \label{enu:The-Herfindahl-index}The Herfindahl index (also known
as Herfindahl–Hirschman Index, HHI, or sometimes HHI-score) is a measure
of the size of firms in relation to the industry they are in and is
an indicator of the amount of competition among them. \href{https://en.wikipedia.org/wiki/Herfindahl\%E2\%80\%93Hirschman_index}{Herfindahl–Hirschman Index, Wikipedia Link}
\item \label{enu:CoinMarketCap}CoinMarketCap is a leading price-tracking
website for cryptoassets in the cryptocurrency space. Its mission
is to make crypto discoverable and efficient globally by empowering
retail users with unbiased, high quality and accurate information
for drawing their own informed conclusions. It was founded in May
2013 by Brandon Chez. \href{https://coinmarketcap.com/about/}{CoinMarketCap, Website Link}
\item \label{enu:Crypto-Ranking}A ranking of cryptocurrencies, including
symbols for the various tokens, by market capitalization is available
on the CoinMarketCap website. We are using the data as of May-25-2022,
when the first version of this article was written. \href{https://coinmarketcap.com}{CoinMarketCap Cryptocurrency Ranking,  Website Link}
\item \label{enu:Cboe-Global-Markets-VIX}Chicago Board Options Exchange
(CBOE) Global Markets revolutionized investing with the creation of
the CBOE Volatility Index® (VIX® Index), the first benchmark index
to measure the market’s expectation of future volatility. The VIX
Index is based on options of the S\&P 500® Index, considered the leading
indicator of the broad U.S. stock market. The VIX Index is recognized
as the world’s premier gauge of U.S. equity market volatility. \href{https://www.cboe.com/tradable_products/vix/}{Chicago Board Options Exchange, VIX Link};
\href{https://en.wikipedia.org/wiki/VIX}{Chicago Board Options Exchange VIX, Wikipedia Link}
\item \label{enu:Blockchain-Bridges}Blockchain bridges work just like the
bridges we know in the physical world. Just as a physical bridge connects
two physical locations, a blockchain bridge connects two blockchain
ecosystems. Bridges facilitate communication between blockchains through
the transfer of information and assets. \href{https://ethereum.org/en/bridges/}{Blockchain Bridges,  Ethereum.Org Website Link}
\item \label{enu:TAO:}Tao or Dao is the natural order of the universe whose
character one's intuition must discern to realize the potential for
individual wisdom. \href{https://en.wikipedia.org/wiki/Tao}{Tao, Wikipedia Link}
\end{enumerate}

\section{References}
\begin{itemize}
\item Aigner, A. A., \& Dhaliwal, G. (2021). Uniswap: Impermanent loss and
risk profile of a liquidity provider. arXiv preprint arXiv:2106.14404.
\item Ang, A., \& Bekaert, G. (2004). How regimes affect asset allocation.
Financial Analysts Journal, 60(2), 86-99.
\item Angeris, G., Kao, H. T., Chiang, R., Noyes, C., \& Chitra, T. (2019).
An analysis of Uniswap markets. arXiv preprint arXiv:1911.03380.
\item Ante, L., Fiedler, I., \& Strehle, E. (2021). The influence of stablecoin
issuances on cryptocurrency markets. Finance Research Letters, 41,
101867.
\item Ao, Z., Horvath, G., \& Zhang, L. (2022). Are decentralized finance
really decentralized? A social network analysis of the Aave protocol
on the Ethereum blockchain. arXiv preprint arXiv:2206.08401.
\item Asness, C. S., Frazzini, A., \& Pedersen, L. H. (2012). Leverage aversion
and risk parity. Financial Analysts Journal, 68(1), 47-59.
\item Auer, R., Frost, J., Gambacorta, L., Monnet, C., Rice, T., \& Shin,
H. S. (2021). Central bank digital currencies: motives, economic implications
and the research frontier. Annual Review of Economics, forthcoming.
\item Agur, I., Ari, A., \& Dell’Ariccia, G. (2022). Designing central bank
digital currencies. Journal of Monetary Economics, 125, 62-79.
\item Baker, A., \& Murphy, R. (2020). Modern monetary theory and the changing
role of tax in society. Social Policy and Society, 19(3), 454-469.
\item Barrdear, J., \& Kumhof, M. (2021). The macroeconomics of central
bank digital currencies. Journal of Economic Dynamics and Control,
104148.
\item Bartoletti, M., Chiang, J. H. Y., \& Lafuente, A. L. (2021). SoK:
lending pools in decentralized finance. In Financial Cryptography
and Data Security. FC 2021 International Workshops: CoDecFin, DeFi,
VOTING, and WTSC, Virtual Event, March 5, 2021, Revised Selected Papers
25 (pp. 553-578). Springer Berlin Heidelberg.
\item Bates, D. S. (2012). US stock market crash risk, 1926–2010. Journal
of Financial Economics, 105(2), 229-259.
\item Belchior, R., Vasconcelos, A., Guerreiro, S., \& Correia, M. (2021).
A survey on blockchain interoperability: Past, present, and future
trends. ACM Computing Surveys (CSUR), 54(8), 1-41.
\item Bianchi, D. (2020). Cryptocurrencies as an asset class? An empirical
assessment. The Journal of Alternative Investments, 23(2), 162-179.
\item Bianchi, D., \& Babiak, M. (2022). On the performance of cryptocurrency
funds. Journal of Banking \& Finance, 138, 106467.
\item Blau, B. M., Griffith, T. G., \& Whitby, R. J. (2021). Inflation and
Bitcoin: A descriptive time-series analysis. Economics Letters, 203,
109848.
\item Blinder, A. S. (1982). The anatomy of double-digit inflation in the
1970s. In Inflation: Causes and effects (pp. 261-282). University
of Chicago Press.
\item Blinder, A. S. (2010). Quantitative easing: entrance and exit strategies.
Federal Reserve Bank of St. Louis Review, 92(6), 465-479.
\item Bodnar, T., \& Schmid, W. (2009). Econometrical analysis of the sample
efficient frontier. The European journal of finance, 15(3), 317-335.
\item Boschen, J. F., \& Weise, C. L. (2003). What starts inflation: evidence
from the OECD countries. Journal of Money, Credit and Banking, 323-349.
\item Bradley, A.C. (1991). Shakespearean Tragedy: Lectures on Hamlet, Othello,
King Lear and Macbeth. London: Penguin. ISBN 978-0-14-053019-3.
\item Briola, A., Vidal-Tomás, D., Wang, Y., \& Aste, T. (2023). Anatomy
of a Stablecoin’s failure: The Terra-Luna case. Finance Research Letters,
51, 103358.
\item Broadie, M. (1993). Computing efficient frontiers using estimated
parameters. Annals of operations research, 45(1), 21-58.
\item Brunner, K., \& Meltzer, A. H. (1971). The uses of money: money in
the theory of an exchange economy. The American Economic Review, 61(5),
784-805.
\item Cai, C. W. (2018). Disruption of financial intermediation by FinTech:
a review on crowdfunding and blockchain. Accounting \& Finance, 58(4),
965-992.
\item Chaplin, M. F. (2001). Water: its importance to life. Biochemistry
and Molecular Biology Education, 29(2), 54-59.
\item Chaves, D., Hsu, J., Li, F., \& Shakernia, O. (2011). Risk parity
portfolio vs. other asset allocation heuristic portfolios. The Journal
of Investing, 20(1), 108-118.
\item Chen, Y. (2018). Blockchain tokens and the potential democratization
of entrepreneurship and innovation. Business horizons, 61(4), 567-575.
\item Choi, S., \& Shin, J. (2022). Bitcoin: An inflation hedge but not
a safe haven. Finance Research Letters, 46, 102379.
\item Chuen, D. L. K., Guo, L., \& Wang, Y. (2017). Cryptocurrency: A new
investment opportunity?. The journal of alternative investments, 20(3),
16-40.
\item Clarke, R., De Silva, H., \& Thorley, S. (2013). Risk parity, maximum
diversification, and minimum variance: An analytic perspective. The
Journal of Portfolio Management, 39(3), 39-53.
\item Cochrane, J. H. (2009). Asset pricing: Revised edition. Princeton
university press.
\item Conlon, T., Corbet, S., \& McGee, R. J. (2021). Inflation and cryptocurrencies
revisited: A time-scale analysis. Economics Letters, 206, 109996.
\item Cosgrove, W. J., \& Loucks, D. P. (2015). Water management: Current
and future challenges and research directions. Water Resources Research,
51(6), 4823-4839.
\item Cousaert, S., Xu, J., \& Matsui, T. (2022). Sok: Yield aggregators
in defi. In 2022 IEEE International Conference on Blockchain and Cryptocurrency
(ICBC) (pp. 1-14). IEEE.
\item Curtis, G. (2004). Modern portfolio theory and behavioral finance. The
Journal of Wealth Management, 7(2), 16-22.
\item Curtis, G. (2002). Modern Portfolio theory and Quantum mechanics.
The Journal of Wealth Management, 5(3), 7-13.
\item Davies, G. (2010). History of money. University of Wales Press.
\item DeLong, J. B. (1997). America's peacetime inflation: the 1970s. In
Reducing inflation: Motivation and strategy (pp. 247-280). University
of Chicago Press.
\item Dimitri, N. (2022). Consensus: Proof of Work, Proof of Stake and Structural
Alternatives. In Enabling the Internet of Value (pp. 29-36). Springer,
Cham.
\item Donmez, A., \& Karaivanov, A. (2022). Transaction fee economics in
the Ethereum blockchain. Economic Inquiry, 60(1), 265-292.
\item Dungey, M., \& Martin, V. L. (2007). Unravelling financial market
linkages during crises. Journal of Applied Econometrics, 22(1), 89-119.
\item Elton, E. J., Gruber, M. J., \& Padberg, M. W. (1978). Simple criteria
for optimal portfolio selection: tracing out the efficient frontier.
The Journal of Finance, 33(1), 296-302.
\item Elton, E. J., \& Gruber, M. J. (1997). Modern portfolio theory, 1950
to date. Journal of banking \& finance, 21(11-12), 1743-1759. 
\item Epstein, G. A. (2019). What's wrong with modern money theory?: A policy
critique. London: Palgrave Macmillan.
\item Fabozzi, F. J., Gupta, F., \& Markowitz, H. M. (2002). The legacy
of modern portfolio theory. The journal of investing, 11(3), 7-22.
\item Fabozzi, F. A., Simonian, J., \& Fabozzi, F. J. (2021). Risk parity:
The democratization of risk in asset allocation. The Journal of Portfolio
Management, 47(5), 41-50.
\item Fama, E. F., \& French, K. R. (1992). The cross‐section of expected
stock returns. the Journal of Finance, 47(2), 427-465.
\item Fama, E. F., \& French, K. R. (2018). Volatility lessons. Financial
Analysts Journal, 74(3), 42-53.
\item Fawley, B. W., \& Neely, C. J. (2013). Four stories of quantitative
easing. Federal Reserve Bank of St. Louis Review, 95(1), 51-88.
\item Flori, A. (2019). News and subjective beliefs: A bayesian approach
to bitcoin investments. Research in International Business and Finance,
50, 336-356.
\item Franks, F. (2000). Water: a matrix of life (Vol. 21). Royal Society
of Chemistry.
\item Fu, S., Wang, Q., Yu, J., \& Chen, S. (2022). FTX Collapse: A Ponzi
Story. arXiv preprint arXiv:2212.09436.
\item Goetzmann, W. N., Ingersoll Jr, J. E., \& Ross, S. A. (2003). High‐water
marks and hedge fund management contracts. The Journal of Finance,
58(4), 1685-1718.
\item Goetzmann, W. N., Brown, S. J., Gruber, M. J., \& Elton, E. J. (2014).
Modern portfolio theory and investment analysis. John Wiley \& Sons,
237.
\item Gonzalez, L. (2020). Blockchain, herding and trust in peer-to-peer
lending. Managerial Finance, 46(6), 815-831.
\item Gorkhali, A., Li, L., \& Shrestha, A. (2020). Blockchain: A literature
review. Journal of Management Analytics, 7(3), 321-343.
\item Grobys, K. (2021). When the blockchain does not block: on hackings
and uncertainty in the cryptocurrency market. Quantitative Finance,
21(8), 1267-1279.
\item Guasoni, P., \& Obłój, J. (2016). The incentives of hedge fund fees
and high‐water marks. Mathematical Finance, 26(2), 269-295.
\item Hanson, R. (2007). Logarithmic markets coring rules for modular combinatorial
information aggregation. The Journal of Prediction Markets, 1(1),
3-15.
\item Hartmann, P., Straetmans, S., \& Vries, C. D. (2004). Asset market
linkages in crisis periods. Review of Economics and Statistics, 86(1),
313-326.
\item Hassija, V., Bansal, G., Chamola, V., Kumar, N., \& Guizani, M. (2020).
Secure lending: Blockchain and prospect theory-based decentralized
credit scoring model. IEEE Transactions on Network Science and Engineering,
7(4), 2566-2575.
\item Healy, A. D., \& Lo, A. W. (2009). Jumping the gates: Using beta-overlay
strategies to hedge liquidity constraints. Journal of investment management,
3(11).
\item Hoang, L. T., \& Baur, D. G. (2021). How stable are stablecoins?.
The European Journal of Finance, 1-17.
\item Hong, K. (2017). Bitcoin as an alternative investment vehicle. Information
Technology and Management, 18(4), 265-275.
\item Hong, H., \& Stein, J. C. (2003). Differences of opinion, short-sales
constraints, and market crashes. The Review of Financial Studies,
16(2), 487-525.
\item Huberts, L. C. (2004). Overlay Speak. The Journal of Investing, 13(3),
22-30.
\item Hughes, L., Dwivedi, Y. K., Misra, S. K., Rana, N. P., Raghavan, V.,
\& Akella, V. (2019). Blockchain research, practice and policy: Applications,
benefits, limitations, emerging research themes and research agenda.
International Journal of Information Management, 49, 114-129.
\item Hull, J. C. (2003). Options futures and other derivatives. Pearson
Education India.
\item Jia, R., \& Yin, S. (2022, November). To EVM or Not to EVM: Blockchain
Compatibility and Network Effects. In Proceedings of the 2022 ACM
CCS Workshop on Decentralized Finance and Security (pp. 23-29).
\item Kashyap, R. (2015). Financial Services, Economic Growth and Well-Being.
Indian Journal of Finance, 9(1), 9-22.
\item Kashyap, R. (2016a). Fighting Uncertainty with Uncertainty. Available
at SSRN 2715424.
\item Kashyap, R. (2016b). Notes on Uncertainty, Unintended Consequences
and Everything Else. Working Paper.
\item Kashyap, R. (2019). For Whom the Bell (Curve) Tolls: A to F, Trade
Your Grade Based on the Net Present Value of Friendships with Financial
Incentives. The Journal of Private Equity, 22(3), 64-81.
\item Kashyap, R. (2020). David vs Goliath (You against the Markets), A
dynamic programming approach to separate the impact and timing of
trading costs. Physica A: Statistical Mechanics and its Applications,
545, 122848.
\item Kashyap, R. (2021). Artificial intelligence: a child’s play. Technological
Forecasting and Social Change, 166, 120555.
\item Kashyap, R. (2021-I). DeFi Security: Turning The Weakest Link Into
The Strongest Attraction. Working Paper.
\item Kashyap, R. (2021-II). Trade Execution: To Trade or Not To Trade.
Working Paper.
\item Kashyap, R. (2021-III). Velocity of Volatility and Variance: Crash
Protection For Crypto-Defi Portfolios. Working Paper.
\item Kashyap, R. (2021-IV). The Risk Parity Line: Moving from The Efficient
Frontier to The Final Frontier of Investments. Working Paper.
\item Kashyap, R. (2021-V). Sharing is Caring: Setting Aside Profits for
The Crypto Community. Working Paper.
\item Kashyap, R. (2021-VI). Raising The Bar for Portfolio Performance Measurement:
The Concentration Risk Indicator. Working Paper.
\item Kashyap, R. (2021-VII). Multichain Expansion and Select Strategic
Initiatives: Building Bridges That Do Not Burn. Working Paper.
\item Kashyap, R. (2021-VIII). The TAO of The DAO: One Small Step for Blockchain
Systems, But A Giant Leap for Mankind. Working Paper.
\item Kashyap, R. (2021-IX). Multi-Asset Fund Flows: Old Money Plus New
Money. Working Paper. 
\item Kashyap, R. (2021-X). Bringing Risk Parity To The Defi Party: A Complete
Solution To The Crypto Asset Management Conundrum. Working Paper.
\item Kashyap, R. (2021-XI). Hedged Mutual Fund Blockchain Protocol: High
Water Marks During Low Market Prices. Working Paper.
\item Kim, A., Trimborn, S., \& Härdle, W. K. (2021). VCRIX—A volatility
index for crypto-currencies. International Review of Financial Analysis,
78, 101915.
\item Klein, T., Thu, H. P., \& Walther, T. (2018). Bitcoin is not the New
Gold–A comparison of volatility, correlation, and portfolio performance.
International Review of Financial Analysis, 59, 105-116.
\item Kosc, K., Sakowski, P., \& Ślepaczuk, R. (2019). Momentum and contrarian
effects on the cryptocurrency market. Physica A: Statistical Mechanics
and its Applications, 523, 691-701.
\item Laidler, D., \& Parkin, M. (1975). Inflation: a survey. The Economic
Journal, 85(340), 741-809.
\item Lee, S., Lee, J., \& Lee, Y. (2022). Dissecting the Terra-LUNA crash:
Evidence from the spillover effect and information flow. Finance Research
Letters, 103590.
\item Lee, S. S., Murashkin, A., Derka, M., \& Gorzny, J. (2022). SoK: Not
Quite Water Under the Bridge: Review of Cross-Chain Bridge Hacks.
arXiv preprint arXiv:2210.16209.
\item Li, X., Jiang, P., Chen, T., Luo, X., \& Wen, Q. (2020). A survey
on the security of blockchain systems. Future Generation Computer
Systems, 107, 841-853.
\item Li, Y., Liu, H., \& Tan, Y. (2022, May). POLYBRIDGE: A Crosschain
Bridge for Heterogeneous Blockchains. In 2022 IEEE International Conference
on Blockchain and Cryptocurrency (ICBC) (pp. 1-2). IEEE.
\item Liu, Y., Tsyvinski, A., \& Wu, X. (2022). Common risk factors in cryptocurrency.
The Journal of Finance, 77(2), 1133-1177.
\item Lucey, B. M., Vigne, S. A., Yarovaya, L., \& Wang, Y. (2022). The
cryptocurrency uncertainty index. Finance Research Letters, 45, 102147.
\item Lyons, R. K., \& Viswanath-Natraj, G. (2023). What keeps stablecoins
stable?. Journal of International Money and Finance, 131, 102777.
\item Madan, D. B., \& Sharaiha, Y. M. (2015). Option overlay strategies.
Quantitative Finance, 15(7), 1175-1190.
\item Mankiw, N. G. (2020, May). A skeptic's guide to modern monetary theory.
In AEA Papers and Proceedings (Vol. 110, pp. 141-44).
\item McLeay, M., Radia, A., \& Thomas, R. (2014). Money in the modern economy:
an introduction. Bank of England Quarterly Bulletin, Q1.
\item Miccolis, J. A. (2012). Management: Putting the “Modern'Back in Modern
Portfolio Theory. Journal of financial Planning.
\item Mohan, V. (2022). Automated market makers and decentralized exchanges:
a DeFi primer. Financial Innovation, 8(1), 20.
\item Mohanty, S. S., Mohanty, O., \& Ivanof, M. (2021). Alpha enhancement
in global equity markets with ESG overlay on factor-based investment
strategies. Risk Management, 23(3), 213-242.
\item Monrat, A. A., Schelén, O., \& Andersson, K. (2019). A survey of blockchain
from the perspectives of applications, challenges, and opportunities.
IEEE Access, 7, 117134-117151.
\item Mulvey, J. M., Ural, C., \& Zhang, Z. (2007). Improving performance
for long-term investors: wide diversification, leverage, and overlay
strategies. Quantitative Finance, 7(2), 175-187.
\item Naeem, M. A., Lucey, B. M., Karim, S., \& Ghafoor, A. (2022). Do financial
volatilities mitigate the risk of cryptocurrency indexes?. Finance
Research Letters, 50, 103206.
\item Nakamoto, S. (2008). Bitcoin: A peer-to-peer electronic cash system.
Decentralized Business Review, 21260.
\item Narayanan, A., \& Clark, J. (2017). Bitcoin's academic pedigree. Communications
of the ACM, 60(12), 36-45.
\item Palley, T. I. (2015). Money, fiscal policy, and interest rates: A
critique of Modern Monetary Theory. Review of Political Economy, 27(1),
1-23.
\item Pahl-Wostl, C. (2008). Requirements for adaptive water management.
In Adaptive and integrated water management (pp. 1-22). Springer,
Berlin, Heidelberg.
\item Patel, S. B., Bhattacharya, P., Tanwar, S., \& Kumar, N. (2020). Kirti:
A blockchain-based credit recommender system for financial institutions.
IEEE Transactions on Network Science and Engineering, 8(2), 1044-1054.
\item Perez‐Quiros, G., \& Timmermann, A. (2000). Firm size and cyclical
variations in stock returns. The Journal of Finance, 55(3), 1229-1262.
\item Qian, E. (2011). Risk parity and diversification. The Journal of Investing,
20(1), 119-127.
\item Reiners, L. (2020). Cryptocurrency and the State: An Unholy Alliance.
S. Cal. Interdisc. LJ, 30, 695.
\item Rhoades, S. A. (1993). The herfindahl-hirschman index. Fed. Res. Bull.,
79, 188.
\item Rogers, P. P., \& Fiering, M. B. (1986). Use of systems analysis in
water management. Water resources research, 22(9S), 146S-158S.
\item Schueffel, P. (2021). DeFi: Decentralized Finance-An Introduction
and Overview. Journal of Innovation Management, 9(3), I-XI.
\item Shahzad, S. J. H., Bouri, E., Roubaud, D., \& Kristoufek, L. (2020).
Safe haven, hedge and diversification for G7 stock markets: Gold versus
bitcoin. Economic Modelling, 87, 212-224.
\item Sharpe, W. F. (1994). The Sharpe Ratio. The Journal of Portfolio Management,
21(1), 49-58.
\item Shleifer, A., \& Vishny, R. W. (1997). The limits of arbitrage. The
Journal of Finance, 52(1), 35-55. 
\item Sipser, M. (2006). Introduction to the Theory of Computation (Vol.
2). Boston: Thomson Course Technology.
\item Slamka, C., Skiera, B., \& Spann, M. (2012). Prediction market performance
and market liquidity: A comparison of automated market makers. IEEE
Transactions on Engineering Management, 60(1), 169-185.
\item Sweeney, J., \& Sweeney, R. J. (1977). Monetary theory and the great
Capitol Hill Baby Sitting Co-op crisis: comment. Journal of Money,
Credit and Banking, 9(1), 86-89.
\item Troster, V., Tiwari, A. K., Shahbaz, M., \& Macedo, D. N. (2019).
Bitcoin returns and risk: A general GARCH and GAS analysis. Finance
Research Letters, 30, 187-193.
\item Uhlig, H. (2022). A Luna-tic Stablecoin Crash (No. w30256). National
Bureau of Economic Research.
\item Van Dijk, M. A. (2011). Is size dead? A review of the size effect
in equity returns. Journal of Banking \& Finance, 35(12), 3263-3274.
\item Veldkamp, L. L. (2005). Slow boom, sudden crash. Journal of Economic
theory, 124(2), 230-257.
\item Wang, S., Yuan, Y., Wang, X., Li, J., Qin, R., \& Wang, F. Y. (2018).
An overview of smart contract: architecture, applications, and future
trends. In 2018 IEEE Intelligent Vehicles Symposium (IV) (pp. 108-113).
IEEE.
\item Wang, H. (2019). VIX and volatility forecasting: A new insight. Physica
A: Statistical Mechanics and its Applications, 533, 121951.
\item Werner, S. M., Perez, D., Gudgeon, L., Klages-Mundt, A., Harz, D.,
\& Knottenbelt, W. J. (2021). Sok: Decentralized finance (defi). arXiv
preprint arXiv:2101.08778.
\item Westall, F., \& Brack, A. (2018). The importance of water for life.
Space Science Reviews, 214(2), 1-23.
\item Wray, L. R. (2015). Modern money theory: A primer on macroeconomics
for sovereign monetary systems. Springer.
\item Xi, D., O’Brien, T. I., \& Irannezhad, E. (2020). Investigating the
investment behaviors in cryptocurrency. The Journal of Alternative
Investments, 23(2), 141-160.
\item Xu, M., Chen, X., \& Kou, G. (2019). A systematic review of blockchain.
Financial Innovation, 5(1), 1-14.
\item Xu, J., Paruch, K., Cousaert, S., \& Feng, Y. (2021). Sok: Decentralized
exchanges (dex) with automated market maker (AMM) protocols. arXiv
preprint arXiv:2103.12732.
\item Xu, J., \& Feng, Y. (2022). Reap the Harvest on Blockchain: A Survey
of Yield Farming Protocols. IEEE Transactions on Network and Service
Management.
\item Yadav, S. P., Agrawal, K. K., Bhati, B. S., Al-Turjman, F., \& Mostarda,
L. (2020). Blockchain-based cryptocurrency regulation: An overview.
Computational Economics, 1-17.
\item Yli-Huumo, J., Ko, D., Choi, S., Park, S., \& Smolander, K. (2016).
Where is current research on blockchain technology?—a systematic review.
PloS one, 11(10), e0163477.
\item Zeng, X., Hao, N., Zheng, J., \& Xu, X. (2019). A consortium blockchain
paradigm on hyperledger-based peer-to-peer lending system. China Communications,
16(8), 38-50.
\item Zetzsche, D. A., Arner, D. W., \& Buckley, R. P. (2020). Decentralized
finance. Journal of Financial Regulation, 6(2), 172-203.
\item Zheng, Z., Xie, S., Dai, H. N., Chen, W., Chen, X., Weng, J., \& Imran,
M. (2020). An overview on smart contracts: Challenges, advances and
platforms. Future Generation Computer Systems, 105, 475-491.
\item Zou, W., Lo, D., Kochhar, P. S., Le, X. B. D., Xia, X., Feng, Y.,
... \& Xu, B. (2019). Smart contract development: Challenges and opportunities.
IEEE Transactions on Software Engineering, 47(10), 2084-2106.
\end{itemize}

\end{document}